\documentclass{aa}  

\usepackage{graphicx}
\usepackage{txfonts}
\usepackage{hyperref}
\def\mobs{M_*^{\mathrm{(obs)}}}

\def\mstar{M_*}
\def\mstartrue{M_*^{\mathrm{(true)}}}
\def\mstartrueein{M_{*,\mathrm{Ein}}^{\mathrm{(true)}}}

\def\detJ{\mathrm{det}J}

\def\msps{M_*^{\mathrm{(sps)}}}

\def\mfive{M_{\mathrm{DM},5}}

\def\tein{\theta_{\mathrm{Ein}}}
\def\tsis{\theta_{\mathrm{Ein}}^{\mathrm{(SIS)}}}
\def\meantsis{<\tsis>}
\def\stdtsis{\sigma(\tsis)}

\def\toneobs{\theta_1^{\mathrm{obs}}}
\def\ttwoobs{\theta_2^{\mathrm{obs}}}
\def\asymm{\xi_{\mathrm{asymm}}}
\def\rmur{r_{\mu_r}}
\def\rmurobs{r_{\mu_r}^{(\mathrm{obs})}}
\def\gammadm{\gamma_{\mathrm{DM}}}
\def\rsnfw{r_s^{(\mathrm{NFW})}}

\def\asps{\alpha_{\mathrm{sps}}}
\def\mumin{\mu_{\mathrm{min}}}
\def\betamax{\beta_{\mathrm{max}}}
\def\betaein{\beta_{\mathrm{Ein}}}

\def\mhalo{M_{200}}

\def\rhalo{r_{200}}
\def\chalo{c_{200}}

\def\cgnfw{c_{200}^{\mathrm{(gNFW)}}}
\def\reff{R_{\mathrm{e}}}

\def\hyperpars{\boldsymbol{\eta}}
\def\indpar{\boldsymbol{\psi}}
\def\indpari{\boldsymbol{\psi}_i}

\def\data{\mathbf{d}}
\def\datai{\mathbf{d}_i}

\def\Sref#1{Section~\ref{#1}\xspace}
\def\Fref#1{Figure~\ref{#1}\xspace}
\def\Tref#1{Table~\ref{#1}\xspace}
\def\Eref#1{Equation~\ref{#1}\xspace}

\def\pr{{\rm P}}

\begin{document}

   \title{Statistical strong lensing. I. Constraints on the inner structure of galaxies from samples of a thousand lenses}
   \titlerunning{Statistical strong lensing. I.}
   \authorrunning{Sonnenfeld \& Cautun}


   \author{Alessandro Sonnenfeld\inst{1}\and
          Marius Cautun\inst{1}\thanks{Marie Sk\l{}odowska-Curie Fellow}
          }

   \institute{Leiden Observatory, Leiden University, Niels Bohrweg 2, 2333 CA Leiden, the Netherlands\\
              \email{sonnenfeld@strw.leidenuniv.nl}
             }

   \date{}

 
  \abstract
    {The number of known strong gravitational lenses is expected to grow substantially in the next few years. The combination of large samples of lenses has the potential to provide strong constraints on the inner structure of galaxies.
}
   {
We investigate the extent to which we can calibrate stellar mass measurements and constrain the average dark matter density profile of galaxies by combining strong lensing data from thousands of lenses.
} 
   {
We generated mock samples of axisymmetric lenses. We assume that, for each lens, we have measurements of two image positions of a strongly lensed background source, as well as magnification information from full surface brightness modelling, and a stellar-population-synthesis-based estimate of the lens stellar mass.
We then fitted models describing the distribution of the stellar population synthesis mismatch parameter $\asps$ (the ratio between the true stellar mass and the stellar-population-synthesis-based estimate) and the dark matter density profile of the population of lenses to an ensemble of 1000 mock lenses.
}
   {
We obtain the average $\asps$, projected dark matter mass, and dark matter density slope with greater precision and accuracy compared with current constraints.
A flexible model and knowledge of the lens detection efficiency as a function of image configuration are required in order to avoid a biased inference.
}
   {
Statistical strong lensing inferences from upcoming surveys provide a way to calibrate stellar mass measurements and to constrain the inner dark matter density profile of massive galaxies.
}

   \maketitle
%

\section{Introduction}

Strong gravitational lensing is one of the few available methods for measuring the masses of galaxies at cosmological distances.
Strong lensing has been used to determine the average density profile of massive galaxies \citep{Koo++06,Aug++10b,Son++13b} and to put constraints on the stellar \citep{Tre++10,Aug++10a,Bar++13,Spi++15,Son++15,SLC15} and dark matter content of these objects \citep{Son++12,O+A18,Sch++19}.

There are two possible approaches to inferring the properties of the mass distribution of galaxies from strong gravitational lensing data.
The first consists of focusing on a selected sample of objects with high-quality data and obtaining as much information as possible from each individual lens. This is the approach adopted, for example, with time-delay lenses for the measurement of cosmological parameters \citep{Suy++17,Mil++20}, and typically involves modelling deep high-resolution images of a lens and combining lensing data with complementary information such as stellar kinematics \citep{Sha++18,Yil++20}.

The second approach consists in combining measurements from a large sample of lenses and inferring the properties of the lens population statistically.
This requires assumptions to be made about the functional form of the distribution of the parameters describing each lens. 
In the simplest case, lenses can be assumed to be homologous systems that are scaled-up versions of each other. Under that assumption, the problem reduces to the determination of a handful of parameters describing the average of the distribution and possible scaling relations between the mass parameters of each lens and some galaxy properties \citep[see, e.g.][]{R+K05,Gri12,ORF14,Sch++14}.
A more general method for inferring the statistical properties of an ensemble of lenses is hierarchical modelling, in which lenses are still assumed to be drawn from a common distribution to be inferred from the data, but where the parameters describing individual objects are allowed to vary independently of each other \citep[see][]{Son++15,Son++19,Bir++20,Sha++21}.
The advantage of a statistical approach to strong lensing inference is that it allows the user to constrain, at a population level, parameters that would otherwise be under-constrained on an individual lens basis.
While large statistics usually implies high precision, not all statistical measurements lead to an accurate result. Any element of complexity in the true distribution of lens properties that is not captured by the model can potentially lead to bias.
The main challenge for a successful statistical strong-lensing measurement is therefore in building a model that is sufficiently flexible to guarantee an accurate answer, yet not too flexible such that it cannot be constrained with strong lensing data alone. This is the problem addressed by this work.

The constraining power of a statistical sample of strong lenses increases with the number of objects. So far, statistical strong-lensing analyses have been carried out on samples of tens of lenses at most, the limiting factor being the availability of spectroscopic data: the redshift of both the lens and the source galaxy is needed to convert angular measurements obtained from the analysis of strongly lensed images into physical measurements of the lens mass.
In the next few years, however, both the number of known lenses and the number of lenses with available spectroscopic observations is expected to grow substantially.
On the one hand, current imaging surveys such as the Hyper Suprime-Cam survey \citep{Aih++18}, the Dark Energy Survey \citep{DES16}, and the Kilo Degree Survey \citep{deJ++15,Kui++15} are leading to the discovery of hundreds of new lenses \citep{Son++18a,Won++18,Pet++19,Jac++19,Cha++20,Son++20,Li++20} and the total number of known lenses is expected to reach approximately $10^5$ with Euclid\footnote{\url{https://www.euclid-ec.org/}} and the Vera Rubin Observatory\footnote{\url{https://www.lsst.org/}} \citep{Col15}.
On the other hand, new spectroscopic facilities such as the Prime Focus Spectrograph\footnote{\url{https://pfs.ipmu.jp/}}, the Dark Energy Spectroscopic Instrument\footnote{\url{https://www.desi.lbl.gov/}}, the 4-metre Multi-Object Spectroscopic Telescope\footnote{\url{https://www.4most.eu/cms/}}, and the Near Infrared Spectrometer and Photometer on board Euclid will offer the opportunity to obtain spectroscopic data for samples of lenses of unprecedented size.

In this study, we investigate the aspects of the mass distribution of galaxies that can be best determined with the statistical combination of strong lensing measurements on a large sample of lenses.
We focus on two properties: the mass-to-light ratio of the stellar component and the inner density profile of the dark matter halo.
Being able to accurately determine the former is crucial for calibrating galaxy stellar mass measurements and therefore obtaining an unbiased account of the baryon cycle in the Universe.
The latter is currently very poorly known and could hold important clues as to the relative importance of baryonic physics processes in galaxy formation and evolution \citep[see][]{Sch++15} or even the nature of dark matter itself.

Statistical strong lensing studies are usually carried out in two steps: at first, each lens is modelled in isolation and its information content is compressed into a handful of parameters summarising the mass distribution of the lens. These inferences on the individual lens parameters are then combined to constrain a model for the lens population.
Here we focus mostly on the second step.

We simulate samples of $1000$ lenses and then try to recover the properties of their population distribution with a Bayesian hierarchical inference method.
We then emulate the lens modelling step: each lens is assumed to be spherical and the observational constraints are compressed into the positions of the two brightest images of a strongly lensed source and the ratio of the radial magnification at these two locations.
This choice allows us to greatly simplify the computational burden of our experiment with respect to a real-world case, while still enabling us to explore the sensitivity of the inference method to a variety of possible systematic effects. These include non-trivial variations in the functional form of the distribution of individual lens parameters, departures of the true dark matter density profile from the family of parameterised models assumed in the fit, and uncertainties in the lens selection function.
We base our simulations both on existing constraints on the structure of strong lenses and on predictions from hydrodynamical simulations.

While it is common to add stellar kinematics constraints to strong lensing data, we do not explore such a possibility here. This is because in order to model stellar kinematics measurements it is necessary to make a series of additional assumptions, for instance on the geometry of the lens and the distribution of the stellar orbits, each of which could introduce a systematic bias that is difficult to quantify.
Instead, we are interested in finding the precision and accuracy with which strong lensing, with the addition of spectroscopic measurements of the lens and source redshift,  can constrain the stellar and dark matter distribution of a large sample of galaxies. 

The structure of this paper is as follows. In \Sref{sect:sl} we introduce the basic concept of strong lensing, including a section describing the aspects of individual lenses that photometric observations can typically constrain. In \Sref{sect:sims} we describe the simulation of the lens population on which our experiments are based. In \Sref{sect:method} we describe the inference method used to analyse the lens sample. In \Sref{sect:results} we show the results of our inference, along with several tests used to quantify the importance of various possible systematic effects. 
In \Sref{sect:discuss} we discuss our results and in \Sref{sect:concl} provide our conclusions.
The Python code used for the simulation and analysis of our lens sample can be found in a dedicated section of a GitHub repository\footnote{\url{https://github.com/astrosonnen/strong_lensing_tools}}.


\section{Strong lensing theory}\label{sect:sl}

\subsection{Basics}

Throughout this work we assume that lenses are (i) isolated, that is they consist of only one galaxy and its dark matter halo, and (ii) circularly symmetric.
Under these assumptions and in the thin lens approximation, which is always valid in the galaxy-scale regime \citep{SEF92}, the lensing properties of a galaxy depend exclusively on its surface mass density projected along the line-of sight, $\Sigma(\theta)$, where $\theta$ is the angular coordinate along an arbitrary axis in the lens plane, also referred to as the image plane, with origin at the lens centre.
A background source at angular position $\beta$ will form images at positions $\theta$ in the lens plane that are solutions of the lens equation:
\begin{equation}
\beta = \theta - \alpha(\theta).
\end{equation}
The variable $\alpha(\theta)$ is the deflection angle and can be calculated from the mass distribution of the lens:
\begin{equation}
\alpha(\theta) = \frac{2}{\theta}\int_0^\theta \frac{\Sigma(\theta')}{\Sigma_{cr}}\theta'd\theta'.
\end{equation}
The integral in the above equation is proportional to the projected mass enclosed within $\theta$, divided by
the critical surface mass density $\Sigma_{cr}$. This is defined as
\begin{equation}
\Sigma_{cr} = \dfrac{c^2D_s}{4\pi G D_d D_{ds}},
\end{equation}
where $c$ is the speed of light and $D_d$, $D_s$, and $D_{ds}$ are the angular diameter distances between the observer and the lens, the observer and the source, and the lens and the source, respectively.
The ratio between the surface mass density of the lens and the critical surface mass density of the lens--source system is defined as the dimensionless surface mass density:
\begin{equation}
\kappa(\theta) \equiv \frac{\Sigma(\theta)}{\Sigma_{cr}}.
\end{equation}

An axisymmetric lens with surface mass density that declines monotonically with distance from the centre can produce either one, two, or three multiple images of the same background source, depending on the source position and on the dimensionless surface mass density profile $\kappa(\theta)$.
Assuming that $\beta > 0$, one image is always produced at $\theta_1 > \tein$, where $\tein$ is the radius of the tangential critical curve or Einstein radius, defined as the solution of the lens equation for $\beta=0$:
\begin{equation}
\tein = \alpha(\tein).
\end{equation}
Depending on the source position, a second image may appear at position $\theta_2$, with $-\tein < \theta_2 < 0$, in which case the source is strongly lensed. A third fainter image may be present at position $\theta_3$ with $\theta_2 < \theta_3 < 0$.

As an illustrative example we consider the case of a power-law lens, with deflection angle given by
\begin{equation}\label{eq:powerlaw}
\alpha^{(\mathrm{PL})}(\theta) = \tein\frac{\theta}{|\theta|}\left(\frac{|\theta|}{\tein}\right)^{2-\gamma}.
\end{equation}
This corresponds to the deflection induced by a spherically symmetric mass distribution with 3D density profile $\rho(r) \propto r^{-\gamma}$.

\begin{figure}
\includegraphics[width=\columnwidth]{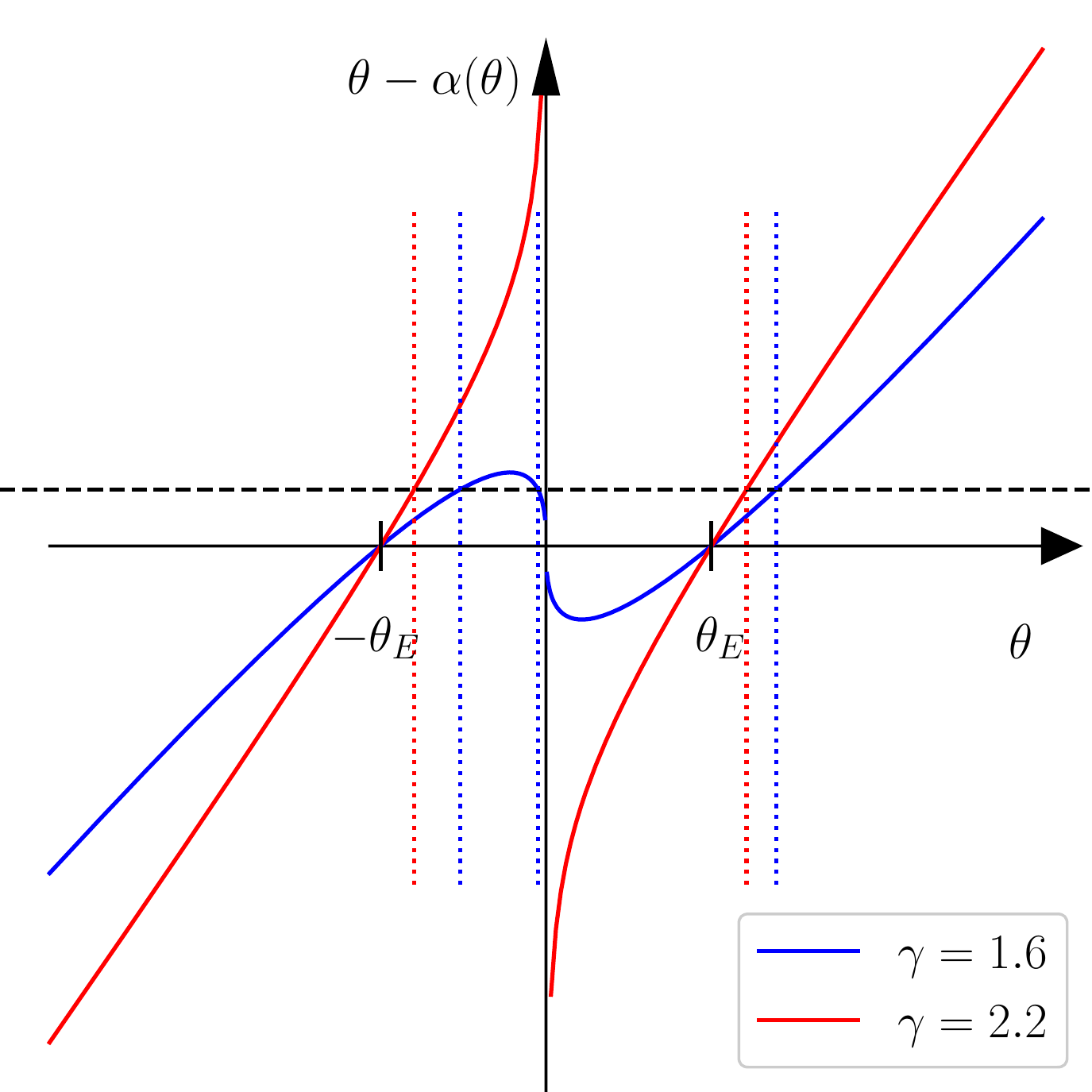}
\caption{
Solutions of the lens equation for axisymmetric power-law lens models. The coloured solid curves show $\theta-\alpha(\theta)$ as a function of $\theta$ for two lenses with the same Einstein radius and different values of the density slope parameter $\gamma$. 
The horizontal dashed line marks the position $\beta$ of a background source. Its images form at solutions of the lens equation, $\beta = \theta -\alpha(\theta)$, indicated by the vertical dotted lines with the colour of the corresponding lens model. For the lens with density profile shallower than isothermal, $\gamma<2$, three images form, while the $\gamma>2$ lens produces only two images.
The slope of the $\theta-\alpha(\theta)$ curve is the inverse of the radial magnification. Stationary points, only visible in the $\gamma<2$ case, correspond to the radial critical curve.
\label{fig:plscheme}
}
\end{figure}

In \Fref{fig:plscheme} we plot the quantity $\theta - \alpha(\theta)$ for two different values of the power-law index $\gamma$ and fixed Einstein radius.
For each lens, images of a background source at position $\beta$ form at values of $\theta$ where the horizontal dashed line intersects the curve, as these points are the solutions to the lens equation.
If $\gamma < 2$, corresponding to a shallower-than-isothermal density profile, the curve $\theta-\alpha(\theta)$ has two stationary points at non-zero values of $\theta$ and, as a result, three images form, provided that $\beta$ is sufficiently small.
These stationary points correspond to the radial critical curve, that is the curve in the image plane where the magnification in the radial direction of an image is infinite.

The radial magnification is given by
\begin{equation}
\mu_r = \left(1 - \frac{d\alpha}{d\theta}\right)^{-1}.
\end{equation}
This is the inverse of the derivative of the function $\theta-\alpha(\theta)$, and is therefore infinite at the stationary points of the function plotted in \Fref{fig:plscheme}.
The total magnification of an image is given by the product between the radial magnification and the magnification in the tangential direction, which is given by
\begin{equation}
\mu_t = \left(1 - \frac{\alpha(\theta)}{\theta}\right)^{-1}.
\end{equation}
As $\theta$ approaches the centre of the lens, the ratio $\alpha/\theta$ becomes very large, and $\mu_t$ tends to zero: for this reason, images close to the centre are typically very faint.

By mapping the radial critical curve to the source plane through the lens equation, we find the position $\beta_r$ of the radial caustic, which delimits the region in the source plane where sources can be strongly lensed: sources with $\beta > \beta_r$ are not strongly lensed into multiple images.
However, not all lenses have a radial critical curve, as can be seen in \Fref{fig:plscheme} in the $\gamma>2$ case. Lenses of this kind always produce two images.
Nevertheless, as the source position moves farther away from the lens, the position $\theta_2$ of the second image gets progressively closer to the centre; both its tangential and radial magnification approach zero, making it invisible.
Regardless of the number of multiple images, in our analysis we only consider the two brighter ones, $\theta_1$ and $\theta_2$, as central images are hardly ever observed in galaxy-scale lenses \citep[see][ for a notable exception]{Sch++19}.

\subsection{Constraints on lens models}

The standard approach to obtaining information on the mass distribution of a lens galaxy involves fitting a lens model to strong lensing data. 
The data consist usually of an image of the lens and the strongly lensed background source, typically made of a large number of pixels.
Modelling a lens requires reproduction of the full surface brightness distribution of the lens and the source.
This is a mature technique \citep[see e.g.][]{W+D03, Suy++06, V+K09, B+A18}, but a time-consuming one, both in terms of human and computational effort.
In order to carry out our experiment within a reasonable time-frame, we emulate the lens modelling process. 
Instead of simulating realistic images of lenses and modelling them, we compress the information content of a lens into a handful of summary observables: the positions and sizes of the two images.
These quantities can be measured robustly with currently available lens modelling tools.
In this section we discuss the properties of a lens that can be recovered with these summary observables.

Two image positions can be used to constrain two degrees of freedom in a lens model. One of these degrees of freedom must be the position $\beta$ of the source, while the other one can be a quantity related to the mass distribution of the lens, for instance the Einstein radius, which can be determined very robustly (i.e. in a model-independent way) when the image configuration is close to symmetric.

When the background source is extended, the two main images have arc-like shapes. If they are well resolved, it is possible to obtain additional constraints on the density profile of the lens by modelling their full surface brightness distribution.
In particular, the width of each arc is proportional to the radial magnification of the lens at its position. While the radial magnification of a single arc is degenerate with the size of the source, which is unknown unless it is a standard ruler, the ratio between the two arc widths is independent of source size and can be used to constrain an additional degree of freedom in the density profile of a lens. 
More precisely, the radial magnification ratio is closely related to the third derivative of the lens potential 
around $\tein$ \citep[see e.g.][]{Son18}.

\begin{figure}
\includegraphics[width=\columnwidth]{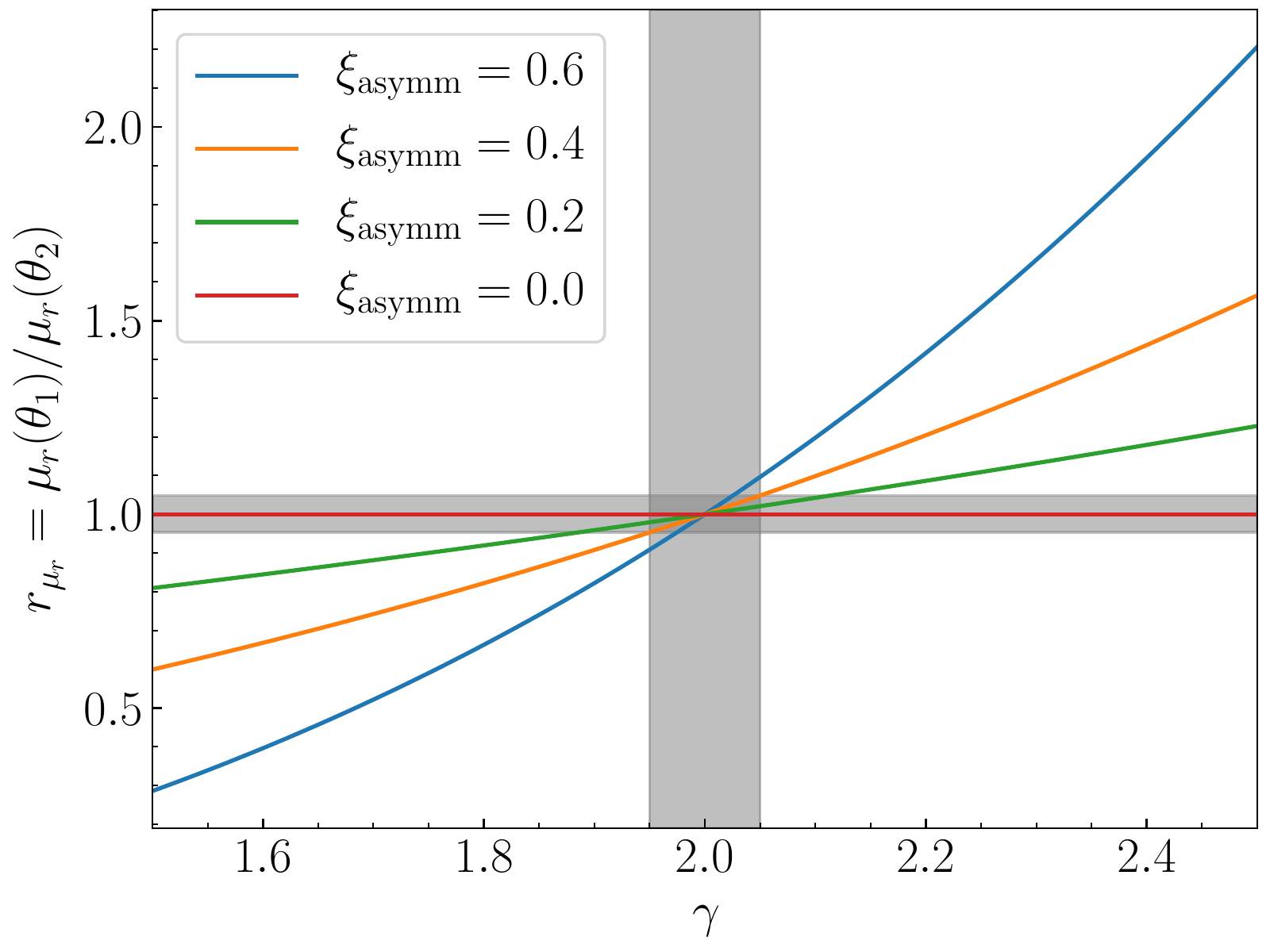}
\caption{
Radial magnification ratio between image 1 and 2 for a lens with a power-law density profile, as a function of the power-law index $\gamma$.
Curves obtained for image configurations with different values of the asymmetry parameter $\asymm$ defined in \Eref{eq:asymm} are shown.
The vertical shaded region indicates the typical uncertainty on the power-law slope, $\Delta\gamma=0.05$, obtained by modelling high-resolution images of strongly lensed extended sources \citep{Sha++21}.
The horizontal shaded region is the uncertainty on $\rmur$ corresponding to an error on the power-law slope of $\Delta\gamma=0.05$ in the case of an image asymmetry $\asymm=0.4$.
\label{fig:rmurpl}
}
\end{figure}

When lens models with a power-law radial dependence of the deflection angle ---described by \Eref{eq:powerlaw}--- are used to fit high-resolution images of strongly lensed extended sources, the slope $\gamma$ of the density profile can be determined from the radial magnification ratio information. 
However, the inferred value of $\gamma$ can be more or less sensitive to the radial magnification ratio, depending on the image configuration.
We illustrate this concept in \Fref{fig:rmurpl}, where we plot the radial magnification ratio between image 1 and 2,
\begin{equation}
\rmur = \frac{\mu_r(\theta_1)}{\mu_r(\theta_2)},
\end{equation}
as a function of the slope $\gamma$, for a few values of the asymmetry parameter $\asymm$, defined as 
\begin{equation}\label{eq:asymm}
\asymm = \frac{\theta_1 + \theta_2}{\theta_1-\theta_2}.
\end{equation} 
For more asymmetric image configurations (larger values of $\asymm$), the curve $\rmur(\gamma)$ is steeper, meaning that a small change in the density slope of the lens model results in a relatively large change in the predicted radial magnification ratio compared to a case in which the image configuration is close to symmetric. If $\rmur$ is determined with a given uncertainty $\Delta\rmur$, the propagated uncertainty on $\gamma$ becomes greater as the value of $\asymm$ decreases.
In the limiting case in which the image consists of a perfect Einstein ring, when the source is at $\beta=0$, the radial magnification ratio between the images is one, independently of the mass model, and therefore it does not have any constraining power.

Based on the above argument, and owing to the popularity of power-law lens models, it is sometimes said that by modelling the full surface brightness distribution of a strongly lensed source it is possible to measure the local slope of the projected density profile at the location of the Einstein radius.
While this statement is true under the assumption that the true density profile of a lens is strictly a power law, it does not hold in general: given a power-law lens model that reproduces the observed image positions and radial magnification ratio, it is always possible to find alternative solutions that fit the data equally well and have different values of the local density slope, because of the mass--sheet degeneracy \citep{FGS85}.
For example, \citet{Bir++20} found that the strong lensing data from the TDCOSMO sample \citep{Mil++20} can be fitted equally well with a pure power-law model or the sum of a scaled-up version of it and a constant mass sheet accounting for up to 10\% of the mass within the Einstein radius.

\subsection{The non-axisymmetric case}

Almost all strong lenses exhibit some departure from axial symmetry. The biggest qualitative difference with respect to the axisymmetric case is that, in the general case, more images of the background source can be formed. Nevertheless, when the source is extended, the image configuration still usually consists  of a main arc and a counter-image.
We can then still summarise the information content of the images of a strongly lensed source with two positions and a radial magnification ratio obtained by comparing the relative widths of the arcs.
As in the axisymmetric case, for a lens with elliptical symmetry, the value of $\rmur$ depends primarily on the third radial derivative of the lens potential at the Einstein radius \citep[compare Equations (16) and (36) of][]{Son18}.
Therefore, the constraining power on the radial mass distribution of such a lens is similar to the axisymmetric case considered in the previous section.
This justifies our choice to treat the lenses as axisymmetric in our experiment.

\section{Simulations}\label{sect:sims}

In this section we describe the procedure that we used to simulate a sample of strong lenses.
We generated strong lenses directly, as opposed to first simulating a population of galaxies and then applying a strong lensing selection.
However, as we explain in section \ref{ssec:sourcepos}, we still take into consideration the fact that some strong lenses are more easily detectable than others when assigning a source to each lens.

Each lens in our sample consists of the sum of a stellar component and a dark matter halo, both concentric and with axial symmetry. 
For the sake of saving computational time, all lenses were taken to be at the same redshift, $z_d=0.4$, and all sources were placed at redshift $z_s=1.5$. 
These values are close to the average of the expected distribution in lens and source redshift from a survey like Euclid \citep{Col15}.
However, our experiment can be generalised to the more realistic case of lenses and sources being distributed in redshift space. 
In the following sections we describe the properties of each element of the lenses and their population distribution in detail.

\subsection{Stellar component}

We describe the stellar mass distribution within each galaxy as a de Vaucouleurs profile:
\begin{equation}
\Sigma_*(R) = \Sigma_0 \exp{\left\{-b\left(\frac{R}{\reff}\right)^{1/4}\right\}},
\end{equation}
where
\begin{equation}
\Sigma_0 = \dfrac{M_* b^{8}}{2\pi\reff^2\Gamma(8)},
\end{equation}
 $\mstar$ is the total stellar mass, $b\simeq=7.669$ is a numerical constant that ensures that the mass enclosed within a radius equal to $R=\reff$ is $\mstar/2$ \citep{C+B99}, and $\Gamma$ is the complete gamma function.

With $\mstartrue$ we indicate the true stellar mass of a galaxy. In addition, we introduce a `stellar population synthesis stellar mass', $\msps$, defined as the stellar mass an observer would measure by fitting a stellar population synthesis model to multi-band photometric data with no errors. 
The quantity $\msps$ is directly accessible from observations, while $\mstartrue$ is not. The former is needed to simulate stellar mass measurements on the lens sample.
The relation between $\msps$ and $\mstartrue$ is described by a parameter $\asps$, which is defined as
\begin{equation}\label{eq:aspsdef}
\mstartrue = \asps\msps.
\end{equation}
We refer to $\asps$ as the stellar population synthesis mismatch parameter.

In past studies, the ratio between the true stellar mass and $\msps$ is usually called the initial mass function (IMF) mismatch parameter, based on the fact that the dominant source of systematic uncertainty when measuring stellar masses photometrically is the choice of the IMF.
However, other choices made during the stellar population synthesis phase, such as priors on the metallicity or the details of the treatment of various evolutionary phases of a stellar population, can also introduce systematic biases in the observed stellar masses. At the precision level that can be reached with large samples of lenses, such systematic errors can be important. We therefore use a more general definition for $\asps$.

We drew values of $\log{\msps}$ from a Gaussian distribution with mean $11.4$ and dispersion $0.3$:
\begin{equation}\label{eq:mspssim}
\log{\msps} \sim \mathcal{N}(11.4, 0.3^2).
\end{equation}
This roughly matches the stellar mass distribution of known samples of strong lenses when measured under the assumption of a Chabrier IMF \citep{Aug++10b,Son++13a,Son++19}.
We then assigned a half-mass radius to each lens, drawn from the following log-Gaussian distribution with a mean that scales linearly with $\log{\msps}$:
\begin{equation}\label{eq:reffsim}
\log{\reff} \sim \mathcal{N}\left(1.0 + 0.8(\log{\msps} - 11.4),\,0.15^2\right),
\end{equation}
where the values of the coefficients were chosen to approximately reproduce the observed stellar mass--size relation of strong lenses from the Sloan Lens ACS Survey \citep[SLACS,][]{Aug++10b}.
Finally, we set $\log{\asps}=0.1$ for all lenses in the sample. This is in the middle of the range of values of the IMF mismatch parameter of strong lenses found in the literature \citep{SLC15,Pos++15,Son++19}.

\subsection{Dark matter halo}\label{ssec:contra}

We drew dark matter halo masses from a log-Gaussian distribution with mean that scales with the stellar mass of a galaxy:
\begin{equation}\label{eq:gaussmhalo}
\log{\mhalo}\sim\mathcal{N}\left(13.0 + 1.5(\log{\msps} - 11.4),\, 0.2^2\right).
\end{equation}
The halo mass $\mhalo$ is defined as the mass enclosed within a spherical shell with mean density equal to 200 times the critical density of the Universe.

We used results obtained from hydrodynamical simulations to define the density profile of each dark matter halo. These consists of modifications to the halo profile found in dark-matter-only simulations, where halos follow a universal profile that is well described by the Navarro, Frenk \& White functional form \citep[NFW;][]{NFW97}:
\begin{equation}\label{eq:gnfw}
\rho(r) = \dfrac{\rho_0}{r/\rsnfw\left(1 + r/\rsnfw\right)^2}.
\end{equation}
For simplicity, in our mocks we imposed a fixed relation between $\rsnfw$ and $\mhalo$. In particular, we set
\begin{equation}\label{eq:scaleradius}
        \rsnfw = \frac{\rhalo}{5},
\end{equation}
where $\rhalo$ is the virial radius, that is the radius of the spherical shell enclosing a mass equal to $\mhalo$. This corresponds to all halos having the same concentration, $c \equiv \rsnfw / \rhalo = 5$. 

The condensation of cold gas at the centre of their halos and the growth of the stellar component leads to deviations \citep[e.g.][]{Blumenthal1986,Gnedin2004} from the NFW profile that are largest in the inner regions of halos, which is the very regime probed by strong lensing. We calculated the changes in the dark matter distribution using the \citet{Cautun2020} relation which has been empirically derived from the \textsc{eagle} and Illustris simulations \citep{Vogelsberger2014,Schaye2015}. The enclosed 3D dark matter mass, $M_{\rm DM}(<r)$, as a function of distance from the halo centre is taken as
{\small
\begin{equation}
   M_{\rm DM}(<r) = (1-f_{\rm bar})M^{\rm(NFW)}(<r) \left[ 0.45 + 0.38 \left( \eta_{\rm bar} + 1.16 \right)^{0.53} \right]
   \label{eq:enclosed_DM_mass} \;,
\end{equation}}
where $f_{\rm bar}$ is the cosmic baryon fraction and $M^{\rm(NFW)}(<r)$ is the enclosed mass of the NFW profile that describe the halo in a dark-matter-only simulation. The $\eta_{\rm bar}(<r)$ parameter describes the level of radial concentration of the baryons with respect to dark matter, and is given by the ratio between the actual enclosed baryonic mass and the expected mass distribution, $f_{\rm bar} M^{\rm(NFW)}(<r)$, assuming baryons follow the same radial profile as the dark matter.

\begin{figure}
\includegraphics[width=\columnwidth]{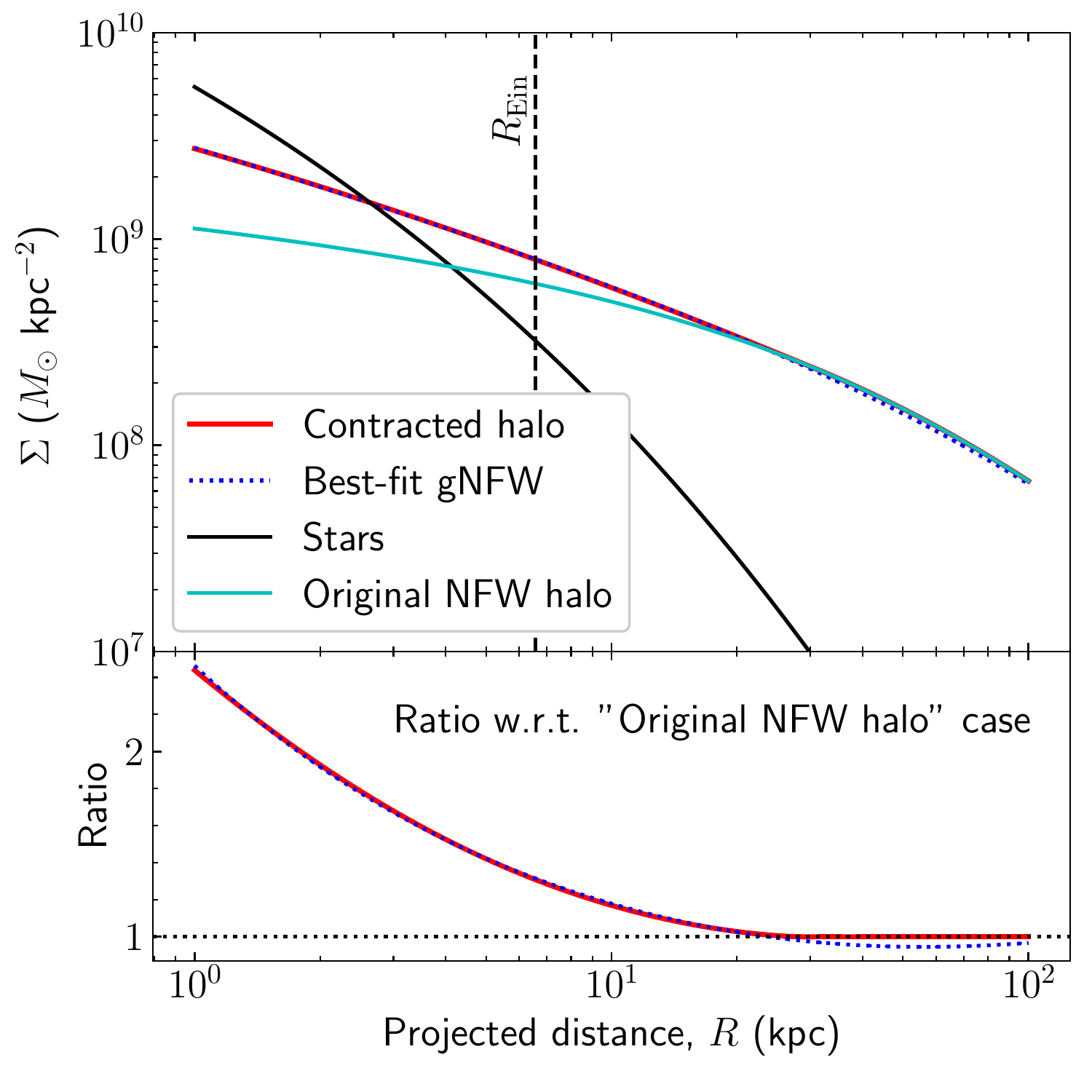}
\caption{
Projected surface mass density of a dark matter halo with mass $\log{\mhalo}=13$, contracted following the procedure described in section \ref{ssec:contra} (magenta line).
Cyan line: Original, pre-contraction dark matter halo described by an NFW profile.
Blue dotted line: gNFW profile fitted to the contracted dark matter halo.
Black line: Stellar component of the lens, consisting of a de Vaucouleurs profile with total mass $\log{\mstar}=11.5$ and half-light radius $\reff=7$~kpc. The values of the halo mass, stellar mass, and half-light radius are close to the median of the distribution of the simulated lens sample.
\label{fig:contra}
}
\end{figure}

In \Fref{fig:contra} we show as an example the projected dark matter density profile obtained with the above procedure for a galaxy with stellar mass $\log{M_*}=11.5$, half-light radius $\reff=7$~kpc, and halo mass $\log{\mhalo}=13$. In the same plot, we show the original NFW density profile of an uncontracted dark matter halo with the same mass (cyan line).

\begin{figure}
\includegraphics[width=\columnwidth]{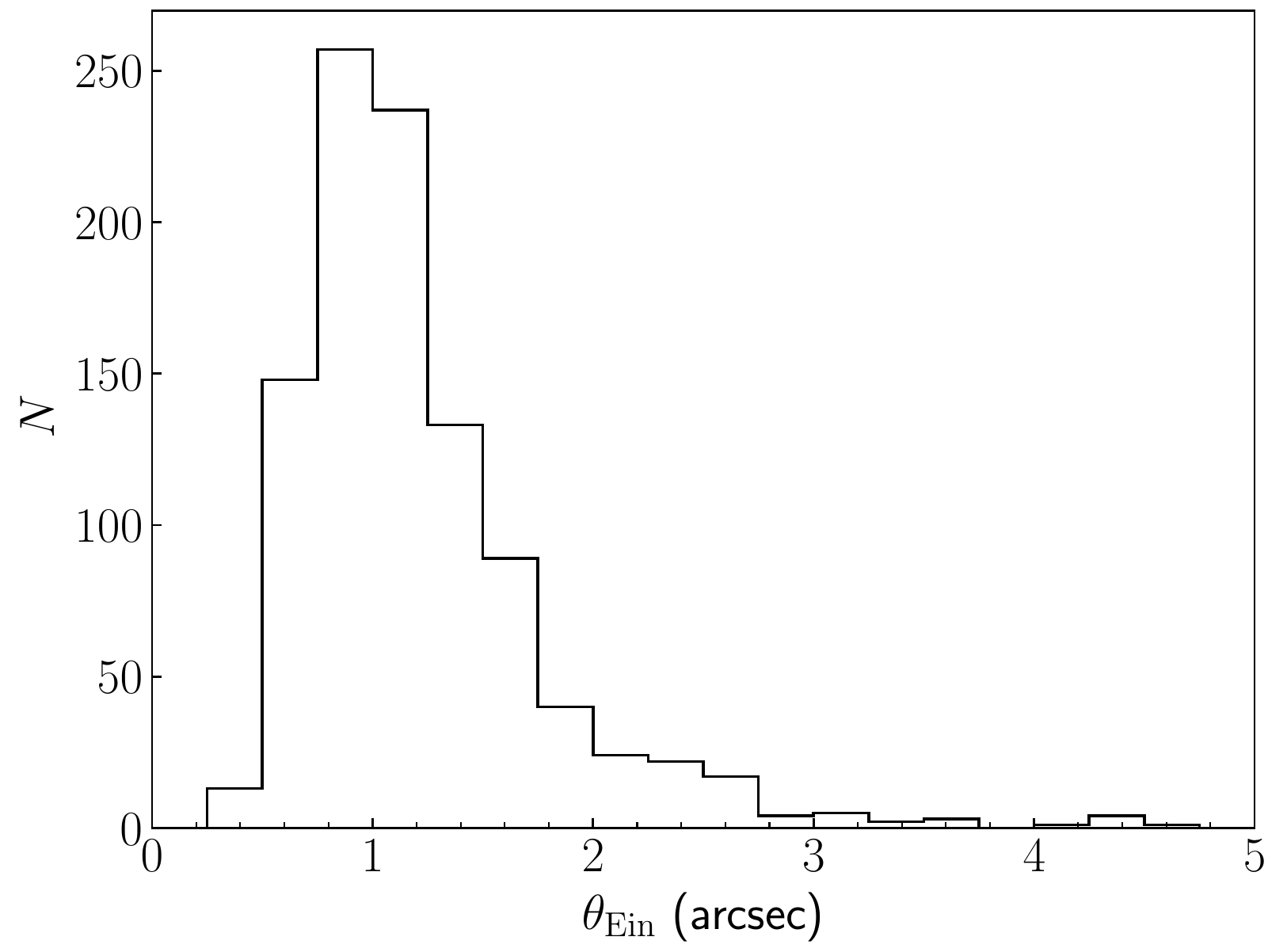}
\caption{
Distribution of the Einstein radii of a sample of 1000 lenses, simulated following the procedure described in \Sref{sect:sims}.
\label{fig:rein}
}
\end{figure}

By applying the prescriptions described so far, we generated a sample of 1000 lenses.
In \Fref{fig:rein} we show the distribution in Einstein radius of the sample.
The bulk of the sample has an Einstein radius in the range $0.5'' < \tein < 2.0''$. This is similar to existing samples of lenses such as the SLACS and the Strong Lensing Legacy Survey \citep[SL2S][]{Son++13a}.

\subsection{Generalised NFW approximation}\label{ssec:gnfw}

The dark matter density profile introduced above is not described by an analytic expression. However, when fitting lensing observations it is convenient to work with analytical models.
A relatively popular choice for the parameterisation of the dark matter density profile of strong lenses is the generalised Navarro Frenk \& White (gNFW) profile:
\begin{equation}\label{eq:gnfw}
\rho(r) = \dfrac{\rho_0}{(r/r_s)^{\gammadm}\left(1 + r/r_s\right)^{3-\gammadm}}.
\end{equation}
A gNFW profile has one additional degree of freedom compared to the standard NFW model: the inner density slope $\gammadm$.
As we explain in \Sref{sect:method}, this is the dark matter density profile that we adopt in the model that we use to fit the simulated data. 

With the goal of understanding how well a gNFW profile can approximate our simulated dark matter halos, we fitted the projected dark matter density of each lens with a gNFW profile.
The fit was done by finding the values of $\gammadm$ and $r_s$ that minimise the difference in projected density on a grid of points logarithmically spaced between $1$ and $30$~kpc, while keeping the value of the halo mass fixed.
The best-fit gNFW model corresponding to the contracted dark matter halo of \Fref{fig:contra} is shown as a red-dotted line in the same plot.
The best-fit values of the inner slope and scale radius are $\gammadm=1.57$ and $r_s=180$~kpc (approximately a factor of $2.3$ larger than the scale radius of the original NFW halo).

\begin{figure*}
\includegraphics[width=\textwidth]{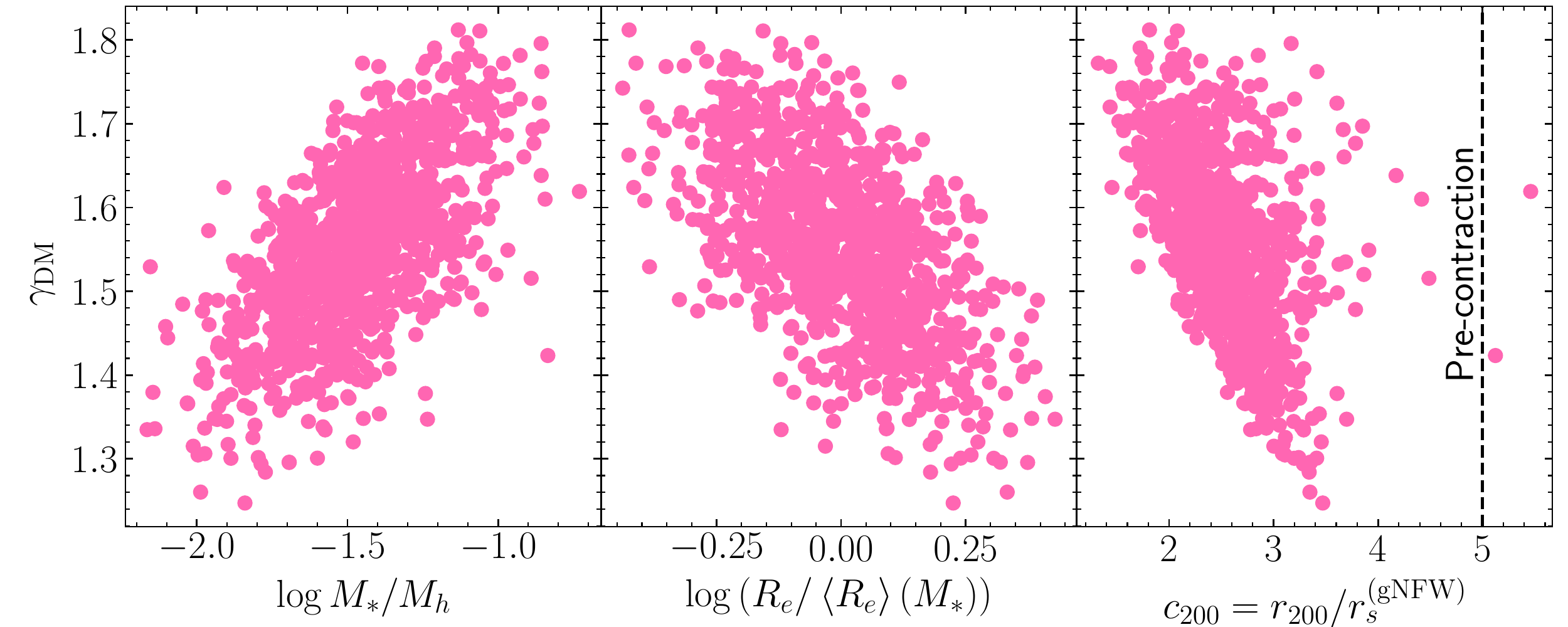}
\caption{
Left panel: Inner density slope of the dark matter halo of the simulated lenses, $\gammadm$, obtained by fitting a gNFW density profile to the projected surface mass density of a lens, as a function of the logarithm of the ratio between the stellar and dark matter halo mass.
Middle panel: $\gammadm$ as a function of the logarithm of the ratio between the stellar half-mass radius and the average half-mass radius of galaxies with the same stellar mass. The latter is given by \Eref{eq:reffsim}.
Right panel: $\gammadm$ as a function of the gNFW concentration parameter, defined as the ratio between the virial radius and the scale radius obtained from the gNFW profile fit. The vertical dashed line marks the value of $\chalo$ adopted for the NFW profile describing the initial (pre-contraction) density profile of the dark matter halo.
\label{fig:slopecorr}
}
\end{figure*}

As the amount of halo contraction depends on the ratio between baryonic and dark matter mass and on the final distribution of the baryons, we expect the inner dark matter slope to be steeper in galaxies with a larger ratio between stellar and halo mass and with a smaller size for a given stellar mass.
Such correlations are indeed observed in our simulated sample, as shown in the left and middle panels of \Fref{fig:slopecorr}.

In the right panel of \Fref{fig:slopecorr}, we plot $\gammadm$ as a function of the concentration parameter $\cgnfw$, defined as the ratio between the virial radius and the scale radius of the best-fit gNFW profile, $r_s^{\mathrm{(gNFW)}}$.
We see that $\gammadm$ is negatively correlated with $\cgnfw$ and that the value of the latter is almost always smaller than $5$, which is the value of the concentration adopted for the initial (pre-contraction) NFW dark matter density profile. 
\subsection{Background source position}\label{ssec:sourcepos}

In a complete sample of strong lenses, the position of the source and that of the lens are not causally related. Therefore, drawing source positions from a uniform distribution in space appears to be an appropriate choice in such a case. However, the farther away the source is from the optical axis, the more asymmetric the image configuration is. Strong lenses with a highly asymmetric image configuration are very difficult to find and model, because the second image tends to be highly de-magnified.

\begin{figure}
\includegraphics[width=\columnwidth]{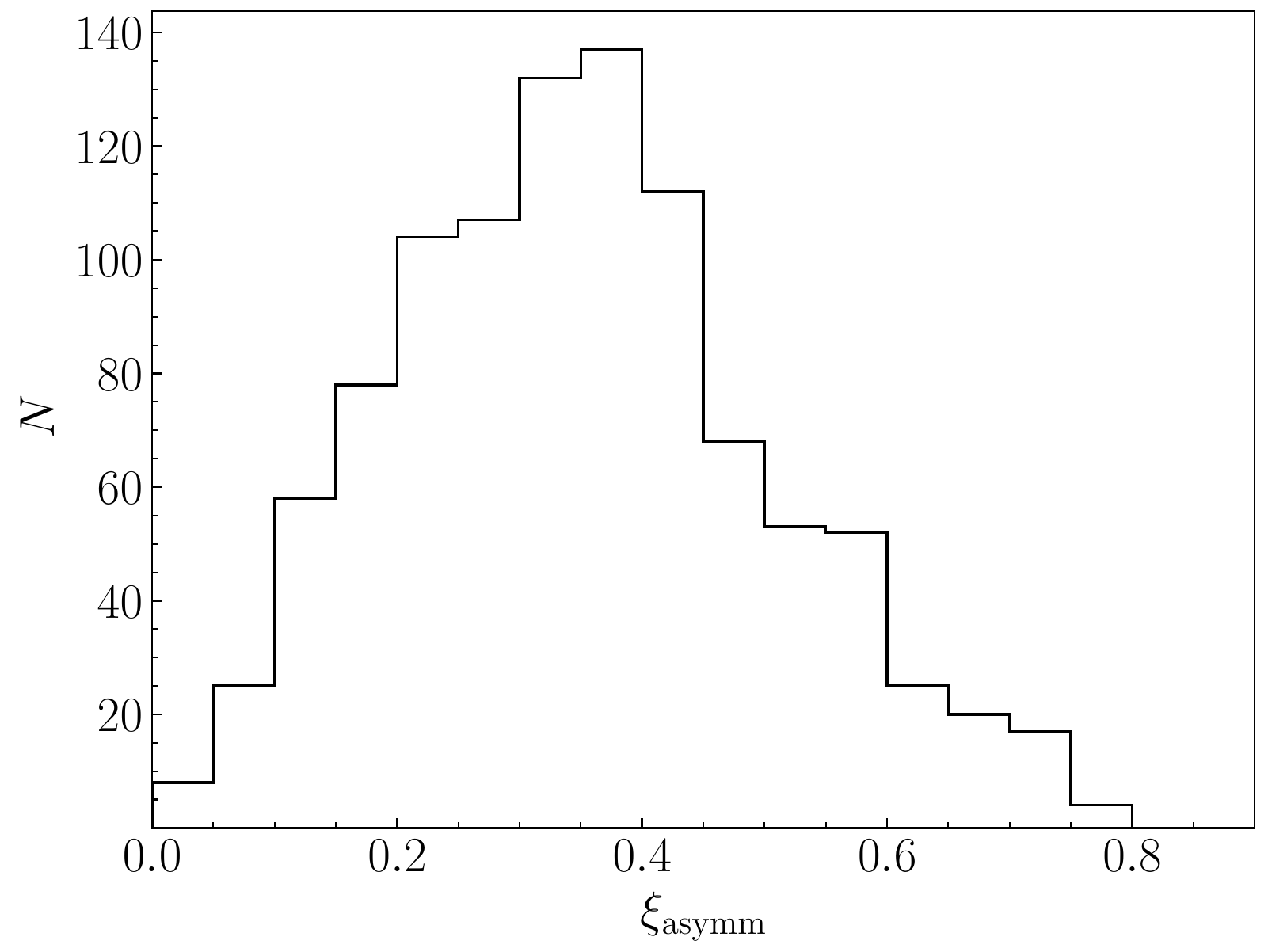}
\caption{
Distribution in the image configuration asymmetry parameter $\asymm$, defined in \Eref{eq:asymm}, of 1000 lenses simulated following the procedure described in \Sref{sect:sims}.
\label{fig:asymm}
}
\end{figure}

We want to exclusively simulate lenses that can realistically be part of a strong lens sample; therefore, we set a limit to how far from the optical axis a source can be for a given lens, based on the corresponding magnification of the second image.
In particular, we found the smallest value of $\beta$ for which the magnification of the second image reaches a minimum allowed value of $\mumin=1$.
We refer to this value as $\betamax$. We then drew a value of $\beta$ from a uniform distribution within a circle of radius $\betamax$:
\begin{equation}
{\rm P}(\beta) \propto \beta\quad{\rm for}\quad\beta<\betamax.
\end{equation}
This is a simplification of what we expect the source position distribution to be in real samples of lenses. The detection efficiency of a lens survey depends not only on the magnification of the second image, but also on the source surface brightness and possibly on the contrast with the lens light.
However, for the purpose of our experiment, the most important feature is the fact that the source position distribution is modified in a non-trivial way from a uniform distribution within the region that is mapped into multiple images.
The resulting distribution in $\asymm$ is shown in \Fref{fig:asymm}.

\subsection{Observational data}

For each lens, we assume that the positions of the two brightest images, $\theta_1$ and $\theta_2$, are measured exactly.
This is a good approximation, because the observational errors on image positions are typically very small (much less than a pixel).
We then assume that the radial magnification ratio between the two images can be measured with a Gaussian error of $\Delta\rmur=0.05$. We model this by adding a Gaussian random error with a mean of zero and a dispersion of 0.05.
We indicate the observed radial magnification ratio as $\rmurobs$ to distinguish it from the true value.
As \Fref{fig:rmurpl} shows, for an image configuration asymmetry of $\asymm=0.4$ (a standard value of this quantity), this translates into an error of $0.05$ on the slope of the density profile of a power-law model, which is the typical uncertainty achieved in lens modelling with current high-resolution data \citep{Sha++21}.
Finally, we added a log-Gaussian noise of $0.15$~dex to the stellar population synthesis-based stellar masses and indicate the resulting values as $\mobs$.
Lens and source redshifts and lens half-light radii are assumed to be known exactly. 
These can typically be determined with very high precision when spectroscopic measurements are available \citep[see e.g.][]{Son++19}.

\section{Inference method}\label{sect:method}

We have a mock sample of $1000$ strong lenses, each with measurements of two image positions, radial magnification ratios, and stellar-population-synthesis-based stellar masses generated as described in \Sref{sect:sims}.
We want to use these data to characterise the distribution of the parameters describing the inner structure of strong lenses. We adopt a Bayesian hierarchical approach for this purpose. 

We assume that the density profile of each lens can be described with a handful of parameters.
We then assume that these parameters are all drawn from a common probability distribution describing the population of lenses. This population distribution is in turn summarised by a small number of high-level parameters, which we refer to as hyper-parameters. Our goal is to constrain the hyper-parameters describing the population.
In the following sections, we describe the different elements of this technique in detail. For past examples of applications of the hierarchical inference formalism to samples of strong lenses, we refer to \citet{Son++15, Son++19}.

\subsection{Individual lens parameters}\label{ssec:individ}

We describe each lens as the sum of a stellar component and a dark matter halo.
We model the stellar component with a de Vaucouleurs profile, which we parameterise by means of the true stellar mass $\mstartrue$ and the half-light radius $\reff$. However, in order to compare our model to the observed stellar mass measurements,  it is also necessary to provide the value of the stellar population synthesis stellar mass, $\msps$. Three parameters then describe the stellar component.

We model the dark matter component with a gNFW profile.
As explained in section \ref{ssec:gnfw}, a gNFW profile has three degrees of freedom. 
However, we are only interested in constraining the average dark matter mass and density slope on the scales probed by strong lensing observations.
We believe that a model with two degrees of freedom in the dark matter density profile is sufficient for that purpose; therefore, we fixed the scale radius to a value of $r_s=100\,{\rm kpc}$ for the sake of reducing the dimensionality of the problem\footnote{One could argue that an NFW profile could be used instead, as it naturally has two degrees of freedom, but varying the scale radius of an NFW profile has only a small effect on the density slope in the inner regions of a dark matter halo.}.
With the goal of working with quantities that are well constrained by our data, we parameterised the dark matter distribution with the projected mass enclosed within $5$~kpc, $\mfive$, and the inner slope $\gammadm$.

Each lens system is then described by a set of six parameters: the true stellar mass, the stellar population synthesis stellar mass, the half-light radius, the projected dark matter mass within $5$~kpc, the inner dark matter density slope, and the position of the source galaxy. We refer to these parameters collectively as
\begin{equation}
\indpar \equiv \{\log{\mstartrue}, \log{\msps}, \reff, \log{\mfive}, \gammadm, \beta \}.
\end{equation}

We point out that, on an individual lens basis, the model is under-constrained, as only five observables per lens are available: the two image positions, the radial magnification ratio, and the observed stellar mass and half-light radius.
We rely on the large sample size and on our statistical model to gain precision on the properties of the lens sample as a whole.

\subsection{Lens population distribution}

The individual lens parameters defined in the previous section are drawn from a probability distribution $\pr(\indpar|\hyperpars)$, where $\hyperpars$ are the hyper-parameters that describe the population of lenses, and  that we want
to infer.
We have the freedom to assert a functional form for this distribution.
Our model must have sufficient flexibility to capture the key features of the lens population that we want to measure. In our case, these features are the average dark matter mass, the average inner dark matter slope, the intrinsic scatter of the dark matter distribution, and the average stellar population synthesis mismatch parameter.
One of the simplest models that can allow us to constrain these properties is the following: 
\begin{equation}\label{eq:popdist}
\begin{split}
\pr(\indpar|\hyperpars) = & \mathcal{S}(\msps,\reff)\,\mathcal{A}\left(\frac{\mstartrue}{\msps}\right)\,\mathcal{H}(\mfive)\times \\ 
& \mathcal{G}(\gammadm)\,\mathcal{B}(\beta|\mstartrue,\reff,\mfive,\gammadm).
\end{split}
\end{equation}
Each term in the above equation describes the distribution of a different property of the lens-source system. 
We now proceed to describe these terms and provide a motivation for each choice.

The term $\mathcal{S}$ in the above equation represents the distribution in the stellar population synthesis stellar mass and half-light radius of the lenses. This term needs to be constrained with the measurements of $\msps$ and $\reff$ of the lens sample. In order to simplify our calculations, we assume that it is known exactly, which means that we fix $\mathcal{S}$ to the product of the two Gaussians of \Eref{eq:mspssim} and \Eref{eq:reffsim}. This is a reasonable assumption, as the distribution in stellar mass and half-light radius of a sample of thousands of galaxies can be determined with high precision \citep[see e.g.][]{SWB19}.

The next term in \Eref{eq:popdist}, labelled $\mathcal{A}$, describes the distribution in the stellar population synthesis mismatch parameter $\asps$, defined in \Eref{eq:aspsdef}.
In principle, this parameter can vary from lens to lens. For simplicity, we assume a single value in our model for the whole population of lenses. Therefore, we write $\mathcal{A}$ as a Dirac delta function:
\begin{equation}
\mathcal{A} = \delta\left(\frac{\mstartrue}{\msps} - \asps\right),
\end{equation}
where $\asps$ is a hyper-parameter of the model in the sense that it describes the distribution of the stellar population synthesis mismatch parameter of the whole population.

The term $\mathcal{H}$ describes the distribution in dark matter mass of the lens sample.
We assume that it has a log-Gaussian functional form, 
\begin{equation}
\mathcal{H}(\mfive) = \frac{1}{\sqrt{2\pi}\sigma_{\mathrm{DM}}}\exp{\left\{-\frac{(\log{\mfive} - \mu_{\mathrm{DM}})^2}{2\sigma_{\mathrm{DM}}^2}\right\}},
\end{equation}
with mean $\mu_{\mathrm{DM}}$ and intrinsic scatter $\sigma_{\mathrm{DM}}$.

The term $\mathcal{G}$ describes the distribution of the inner dark matter slope.
We assume a Gaussian distribution for it, truncated for $\gammadm < 0.8$ and $\gammadm > 1.8$:
\begin{equation}\label{eq:gammadist}
\mathcal{G}(\gammadm) = \frac{A_\gamma}{\sqrt{2\pi}\sigma_\gamma}\exp{\left\{-\frac{(\gammadm - \mu_\gamma)^2}{2\sigma_\gamma^2}\right\}}.
\end{equation}
The coefficient $A_\gamma$ is a normalisation constant that ensures that the integral over $\gammadm$ of $\mathcal{G}$ on its support, $(0.8, 1.8)$, is one.

The motivation for the upper bound on $\gammadm$ is that we assert that the density profile of the dark matter halo must be shallower than that of the total matter. As typical lenses have a total density profile close to isothermal, $\rho(r) \propto r^{-2}$ \citep{Koo++06}, this is achieved by truncating the distribution of the dark matter slope at $\gammadm=1.8$. The lower bound at $\gammadm=0.8$ is imposed purely to speed up computations by reducing the volume of the parameter space. We verified that the results do not change by modifying the value of the lower bound.

Finally, the term $\mathcal{B}$ describes the distribution in the source position $\beta$.
As explained in section \ref{ssec:sourcepos}, this is directly related to the selection function of the strong lens sample: at fixed lens density profile, the position of the source determines the brightness of the multiple images and therefore their detectability.
For simplicity, we assume that the source position distribution, and implicitly also the lens sample detection efficiency, are known exactly. We discuss the impact of this assumption in section \ref{ssec:betaprior}.
Given the procedure that was used to assign source positions to the mock lenses, the term $\mathcal{B}$ is therefore
\begin{equation}\label{eq:betadist}
\mathcal{B}(\beta|\mstartrue,\reff,\mhalo,\gammadm) = \left\{\begin{array}{ll} \dfrac{2\beta}{\betamax^2} & \rm{if}\,0 < \beta < \betamax \\
& \\
0 & \rm{elsewhere}\end{array}\right. .
\end{equation}
In other words, the source position distribution is uniform within a circle of radius $\betamax$, where $\betamax$ is the smallest\footnote{The magnification of the second image is not necessarily a monotonic function of $\beta$} value of $\beta$ for which the magnification of the second image is equal to $\mumin=1$. The value of $\betamax$ depends in turn on the lens structural parameters $\mstar$, $\reff$, $\mhalo$ and $\gammadm$.

We refer to the model described so far as the `base model', to distinguish it from more complex models that we introduce in the following section.
We stress that this model does not correspond to the true mass distribution of the simulated sample of lenses for any value of its hyper-parameters because of the differences in the description of the dark matter density profile (both on a single lens basis and in terms of the population distribution).
This was a deliberate choice, the aim being to reproduce the conditions of an inference on real data, in which any model that is fitted is inevitably only an approximation of the truth.

\subsection{Inference technique}\label{ssec:inferencetech}

We need to estimate the posterior probability distribution function of the model hyper-parameters given the data, $\pr(\hyperpars|\data)$. From Bayes theorem, this is proportional to the product of the prior probability of the hyper-parameters, $\pr(\hyperpars)$, multiplied by the likelihood of observing the data given the hyper-parameters, $\pr(\data|\hyperpars)$:
\begin{equation}\label{eq:bayes}
\pr(\hyperpars|\data) \propto \pr(\hyperpars)\pr(\data|\hyperpars).
\end{equation}
As measurements performed on the different lenses are independent of each other, the likelihood can be written as the following product over the lenses:
\begin{equation}\label{eq:likelihood}
\pr(\data|\hyperpars) = \prod_i \pr(\datai|\hyperpars),
\end{equation}
where $\datai$ indicates the observational data of the $i-$th lens. These consist of the two image positions $(\toneobs,\ttwoobs)$, the radial magnification ratio $\rmurobs$, the observed (stellar population model-dependent) stellar mass $\mobs$ , and related uncertainties.

In addition to the hyper-parameters, these data depend on the parameters describing each lens, $\indpari$. In order to evaluate $\pr(\datai|
\hyperpars)$,  it is therefore necessary to consider all possible values taken by the individual lens parameters $\indpari$, that is to marginalise over them:
\begin{equation}\label{eq:fullintegral}
\pr(\datai|\hyperpars) = \int d\indpari \pr(\datai|\indpari,\hyperpars) \pr(\indpari|\hyperpars).
\end{equation}
Formally, $\indpari$ is a six-dimensional variable. Of the integrals over these dimensions, the one over $\reff$ is a trivial one, as we assume that the half-light radius is measured exactly (the likelihood in the half-light radius is a Dirac delta function centred on the true value).
Consequently, at fixed true stellar mass $\mstartrue$, the integral over $\log{\msps}$ returns the value of the integrand evaluated at $\msps = \mstartrue/\asps$.
In other words, the value of the hyper-parameter $\asps$ and the value of $\mstartrue$ determine $\msps$ exactly.
\Eref{eq:fullintegral} subsequently becomes the following four-dimensional integral:
\begin{equation}\label{eq:4dintegral}
\begin{split}
\pr(\datai|\hyperpars) = & \int d\gammadm \int d\log{\mfive} \int d\log{\mstartrue} \int d\beta \\ & \pr\left(\datai|\mstartrue,\asps,\reff,\mfive,\gammadm,\beta\right) \\
 & \pr\left(\mstartrue,\reff,\mfive,\gammadm,\beta|\hyperpars\right),
\end{split}
\end{equation}
where we omit the subscript $i$ on the lens parameter variables for the sake of keeping the notation compact.
Because the two image positions are measured exactly, two of these integrals are integrals over Dirac delta functions, which can be computed analytically. As we show in Appendix~\ref{sect:appendixa}, integrating over $\beta$ and $\log{\mstartrue}$ we obtain 
\begin{equation}\label{eq:2dintegral}
\begin{split}
\pr(\datai|\hyperpars) = & \int d\gammadm \int d\log{\mfive} \left\lvert\detJ\right\rvert_{(\mstartrue,\beta)=(\mstartrueein,\betaein)}\\
& \pr(\rmurobs|\gammadm,\mfive,\reff,\mstartrueein,\betaein) \\
& \pr\left(\mobs|\mstartrueein,\asps\right) \\
& \pr\left(\mstartrueein,\reff,\mfive,\gammadm,\betaein|\hyperpars\right).
\end{split}
\end{equation}
In the above equation, $\mstartrueein$ and $\betaein$ are the values of the stellar mass and source position that, for a given combination of the parameters $(\mfive,\gammadm)$, are needed to reproduce the two image positions, $\toneobs$ and $\ttwoobs$. 
The term $\detJ$ is the Jacobian determinant corresponding to the following variable change,
\begin{equation}
(\log{\mstartrue},\beta) \rightarrow (\theta_1,\theta_2),
\end{equation}
evaluated at $\mstartrueein$ and $\betaein$. The Jacobian determinant is also a function of $\mfive$ and $\gammadm$.

\Eref{eq:2dintegral} is a two-dimensional integral. While this is much more tractable than that of \Eref{eq:4dintegral}, it still needs to be evaluated numerically. 
The precision requirement on the calculation of these integrals is very high: as the likelihood of the hyper-parameters given the data, \Eref{eq:likelihood}, is the product of a thousand such terms, a small systematic error in the calculation of \Eref{eq:2dintegral} can introduce large biases in the posterior probability.
For instance, a $0.1\%$ error on each $\pr(\datai|\hyperpars)$ term translates into a factor $2.7$ error on the product of $1000$ such terms.

We calculated the integrals of \Eref{eq:2dintegral} via spline integration. 
We first defined a two-dimensional grid in the $(\gammadm,\log{\mfive})$ parameter space. We then evaluated the integrand function at each point on the grid. This required calculation of the values of $\mstartrueein$, $\betaein$, and $\detJ$ for each value of $(\gammadm,\log{\mfive}$), which was done only once per lens at the beginning of the analysis. Subsequently, for each value of $\gammadm$ on the grid, we approximated the integrand function with a third-order polynomial spline in $\log{\mfive}$ and used it to integrate over $\log{\mfive}$. Finally, we repeated this procedure over the $\gammadm$ variable.

We sampled the posterior probability distribution of the hyper-parameters given the data using {\sc emcee} \citep{For++13}, the Python implementation of the affine-invariant sampling method introduced by \citet{G+W10}.
We assumed flat priors over finite intervals for all hyper-parameters, as described in the first column of \Tref{tab:base_inference}.
We verified that our inference method is accurate by applying it to a mock sample of lenses generated from the same model family assumed in this section. 
We also verified that the inference is converged with respect to the resolution of the $(\gammadm,\log{\mfive})$ grid used for the computation of the integrals of \Eref{eq:2dintegral}.

\section{Results}\label{sect:results}

\begin{figure*}
\includegraphics[width=\textwidth]{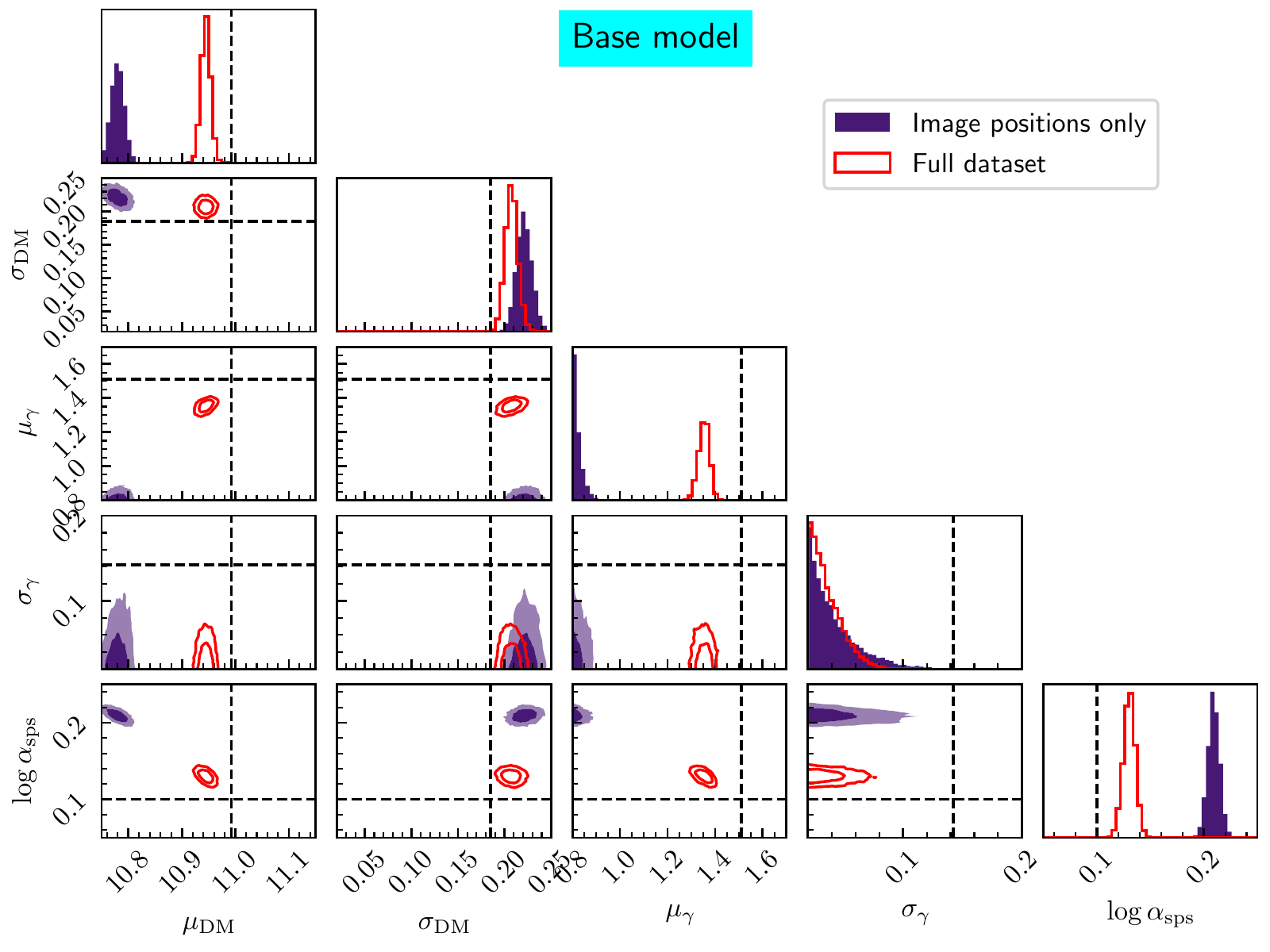}
\caption{
Posterior probability distribution of the hyper-parameters of the model described in \Sref{sect:method}, dubbed the `base model', given the mock data of a sample of 1000 lenses generated with the procedure described in \Sref{sect:sims}.
Red lines show the fit to the whole dataset (image positions and radial magnification ratios). Filled contours show the fit to image position only.
Contour levels correspond to 68\% and 95\% enclosed probability regions.
Dashed lines indicate the true values of the hyper-parameters, which are defined by fitting each model directly to the distribution of $\log{\mhalo}$, $\gammadm$, and $\log{\asps}$ of the mock sample.
\label{fig:cp1}
}
\end{figure*}

\Fref{fig:cp1}  shows the posterior probability distribution of the hyper-parameters of the model described in \Sref{sect:method} given the simulated data described in \Sref{sect:sims}. The median and the 16th and 84th percentiles of the marginal posterior of each hyper-parameter are reported in the second column of \Tref{tab:base_inference}.

Both in \Fref{fig:cp1} and in \Tref{tab:base_inference} we report the true values of the hyper-parameters. The true values of the hyper-parameters describing the distribution in the dark matter mass and slope were defined by fitting our base model directly to the individual values of $\mfive$ and $\gammadm$ of the lenses. The inner slope $\gammadm$ was defined by fitting a gNFW profile with $r_s=100$~kpc and the true value of $\mfive$ to the projected dark matter mass in the range $1-30$~kpc. This procedure is different from the one adopted in section \ref{ssec:gnfw}; therefore, the resulting values of $\gammadm$ are slightly different from those shown in \Fref{fig:slopecorr}.
Visual inspection suggests that the distributions in $\mfive$ and $\gammadm$ of the sample appear qualitatively close to Gaussian.

\begin{table*}
\caption{Inference on the hyper-parameters of the base model given mock data from a sample of 1000 strong lenses. 
Column (2): true values of the hyper-parameters. For the hyper-parameters relative to the inner dark matter slope, these are defined by fitting the model directly to the distribution of $\mfive$ and $\gammadm$. 
Column (3): priors on the hyper-parameters.
Columns (4)-(5): median, 16th and 84th percentile of the marginal posterior probability distribution of each hyper-parameter given the full dataset (image positions and radial magnification ratios) and image position data only.}
\label{tab:base_inference}
\begin{tabular}{lccccl}
\hline
\hline
Parameter & Truth & Prior & Full data & Image pos. only & Description \\
\hline
$\mu_{\mathrm{DM}}$ & $10.99$ & $U(10.00,12.00)$ & $10.945_{-0.009}^{+0.009}$ & $10.780_{-0.013}^{+0.013}$ & Mean $\log{\mfive}$ \\
$\sigma_{\mathrm{DM}}$ & $0.19$ & $U(0.02,0.50)$ & $0.207_{-0.007}^{+0.007}$ & $0.221_{-0.008}^{+0.009}$ & Intrinsic scatter in $\log{\mfive}$ \\
$\mu_{\gamma}$ & $1.51$ & $U(0.80,1.80)$ & $1.35_{-0.02}^{+0.02}$ & $0.816_{-0.012}^{+0.026}$ & Mean $\gammadm$ \\
$\sigma_{\gamma}$ & $0.14$ & $U(0.02,0.50)$ & $0.033_{-0.009}^{+0.016}$ & $0.036_{-0.012}^{+0.027}$ & Intrinsic scatter in $\gammadm$ \\
$\log{\alpha_{\mathrm{sps}}}$ & $0.10$ & $U(0.00,0.25)$ & $0.130_{-0.006}^{+0.005}$ & $0.209_{-0.006}^{+0.006}$ & Log of the stellar population synthesis mismatch \\
 & & & & & parameter \\\end{tabular}
\end{table*}

The inference is very precise: The uncertainties on the hyper-parameters are very small compared to current constraints on the dark matter density profile and stellar IMF of strong lenses.
However, it is not accurate: The true values of all hyper-parameters lie outside of the 95\% credible region of the posterior probability distribution. 

\subsection{Extending the model}\label{ssec:extended}

When fitting the base model introduced in \Sref{sect:method} to our mock sample of lenses, we obtain an inference with high precision but poor accuracy. In other words, we are in a systematic-errors-dominated regime.
We can try to gain accuracy by adding flexibility to the model.
The base model does not allow for correlations between the dark matter parameters and any other property of the lenses. 
Such correlations are present in the mock sample, as shown in \Fref{fig:slopecorr}, and more generally it is reasonable to believe that the distribution of stars in a galaxy is linked to the distribution of dark matter.

We then generalise the base model by modifying the mean parameter of the $\mfive$ and $\gammadm$ distributions as follows:
\begin{eqnarray}
\mu_{\mathrm{DM}} & = & \mu_{\mathrm{DM},0} + \beta_{\mathrm{DM}}(\log{\msps} - 11.4) + \nonumber \\
& & \xi_{\mathrm{DM}}(\log{\reff} - \mu_R(\msps)) \\
\mu_\gamma & = & \mu_{\gamma,0} + \beta_\gamma(\log{\msps} - 11.4) + \nonumber \\
& & \xi_\gamma(\log{\reff} - \mu_R(\msps)),
\end{eqnarray}
where $\mu_R(\msps)$ is the average value of $\log{\reff}$ of lenses with stellar-population-synthesis-derived stellar mass $\msps$.
We introduced four new parameters: $\beta_{\mathrm{DM}}$ and $\beta_\gamma$ describe the correlation between $\mfive$ and $\gammadm$ and the stellar mass, while $\xi_{\mathrm{DM}}$ and $\xi_\gamma$ describe correlations between $\mfive$ and $\gammadm$ and the ratio between the size of a galaxy and the average size of galaxies of the same stellar mass.
In principle, we could also add an explicit correlation between $\mfive$ and $\gammadm$, but we chose not to for the sake of simplicity.
All other aspects of the model are kept as in the base model. We refer to this as the extended model.

We first measured the true values of the new set of hyper-parameters related to the inner dark matter slope by fitting the extended model directly to the distribution of $\gammadm$. These are reported in \Tref{tab:extended_inference} and shown in \Fref{fig:cp1} as black dashed lines.

As the stellar mass increases, the projected dark matter mass within 5~kpc also increases, albeit in a sublinear way: $\beta_{\mathrm{DM}} = 0.60$. Conversely, the inner dark matter slope decreases: $\beta_\gamma = -0.41$.
At fixed stellar mass, galaxies with a larger half-light radius have both a smaller dark matter mass and a shallower dark matter slope: $\xi_{\mathrm{DM}}=-0.21$ and $\xi_\gamma=-0.34$.
The values of $\mu_\gamma$ and $\sigma_\gamma$ are also modified with respect to those obtained when fitting the base model. In particular, the intrinsic scatter is much smaller: this is because part of the scatter observed in the context of the base model can be accounted for by correlations with $\msps$ and $\reff$.

\begin{figure*}
\includegraphics[width=\textwidth]{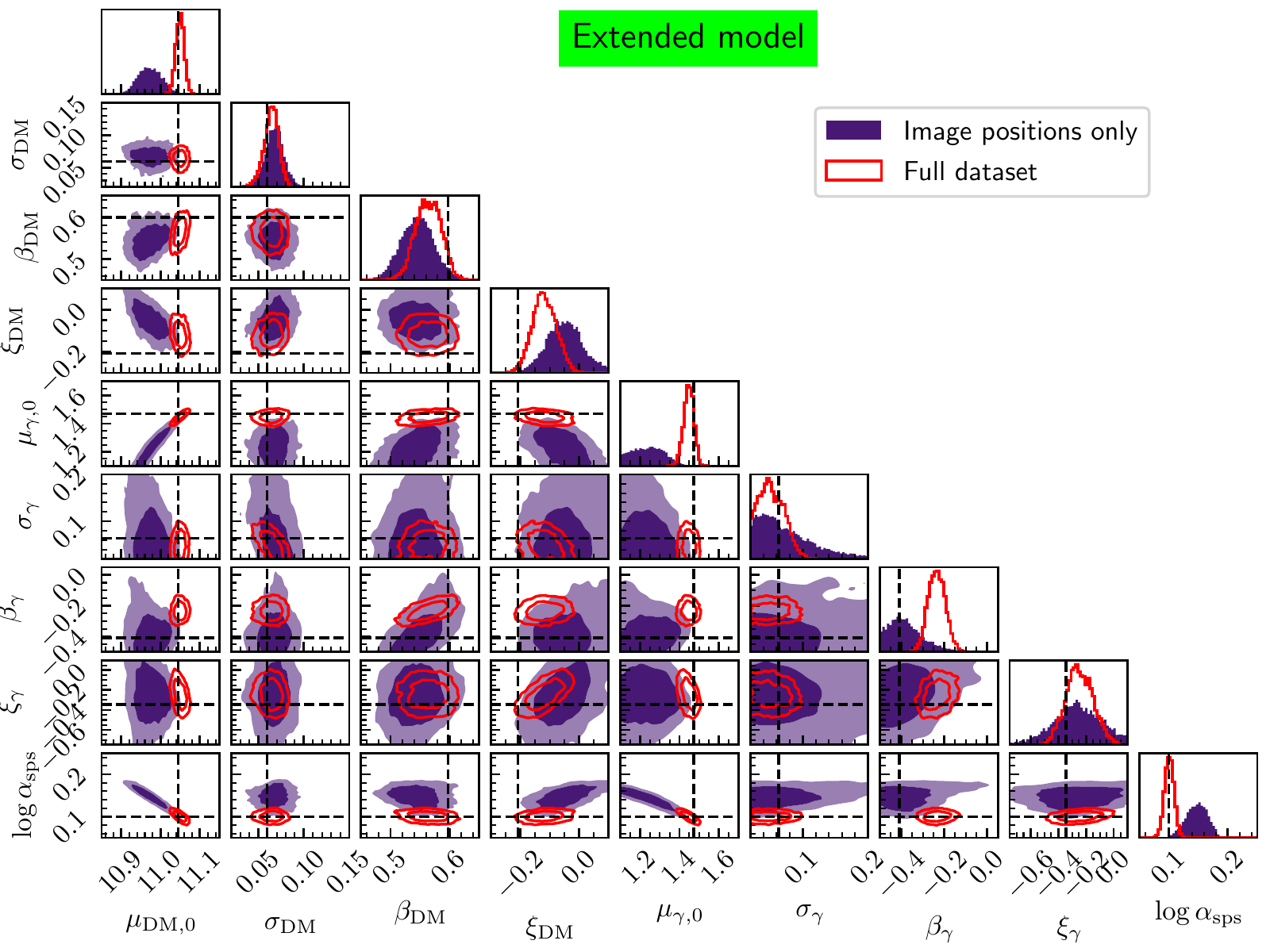}
\caption{
Posterior probability distribution of the hyper-parameters of the extended model introduced in section \ref{ssec:extended} given the mock data of a sample of 1000 lenses.
Red lines show the fit to the whole dataset (image positions and radial magnification ratios). Filled contours show the fit to image position only.
Contour levels correspond to 68\% and 95\% enclosed probability regions.
Dashed lines indicate the true values of the hyper-parameters, which are defined by fitting the each model directly to the distribution of $\log{\mhalo}$, $\gammadm$, and $\log{\asps}$ of the mock sample.
\label{fig:cp2}
}
\end{figure*}

In \Fref{fig:cp2} we show the posterior probability distribution of the hyper-parameters of the extended model given the mock data, in red contours.
The inferred marginal posterior probability distribution of each parameter is summarised in \Tref{tab:extended_inference}.
The extended model allows for a much more accurate inference of all hyper-parameters compared to the base model. All true values are recovered, with the exception of the parameter describing the stellar mass dependence of the dark matter slope, $\beta_\gamma$.

\begin{table*}
\caption{Inference on the hyper-parameters of the extended model given mock data from a sample of 1000 strong lenses. 
Column (2): true values of the hyper-parameters. For the hyper-parameters relative to the inner dark matter slope, these are defined by fitting the model directly to the distribution of $\mfive$ and $\gammadm$. 
Column (3): priors on the hyper-parameters.
Columns (4)-(5): median, 16th and 84th percentile of the marginal posterior probability distribution of each hyper-parameter given the full dataset (image positions and radial magnification ratios) and image position data only.}
\label{tab:extended_inference}
\begin{tabular}{lccccl}
\hline
\hline
Parameter & Truth & Prior & Full data & Image pos. only & Description \\
\hline
$\mu_{\mathrm{DM}, 0}$ & $11.05$ & $U(10.00,12.00)$ & $11.050_{-0.011}^{+0.010}$ & $10.97_{-0.03}^{+0.03}$ & Mean $\log{\mfive}$ at $\log{\msps}=11.4$ and average size \\
$\beta_{\mathrm{DM}}$ & $0.60$ & $U(0.00,3.00)$ & $0.56_{-0.02}^{+0.02}$ & $0.55_{-0.03}^{+0.03}$ & Dependence of $\log{\mfive}$ on $\msps$ \\
$\xi_{\mathrm{DM}}$ & $-0.21$ & $U(-1.00,1.00)$ & $-0.12_{-0.04}^{+0.04}$ & $-0.05_{-0.06}^{+0.06}$ & Dependence of $\log{\mfive}$ on galaxy size \\
$\sigma_{\mathrm{DM}}$ & $0.06$ & $U(0.02,0.50)$ & $0.065_{-0.008}^{+0.007}$ & $0.068_{-0.011}^{+0.010}$ & Intrinsic scatter in $\log{\mfive}$ \\
$\mu_{\gamma,0}$ & $1.47$ & $U(0.80,1.80)$ & $1.45_{-0.03}^{+0.02}$ & $1.21_{-0.12}^{+0.10}$ & Mean $\gammadm$ at $\log{\msps}=11.4$ and average size \\
$\beta_{\gamma}$ & $-0.41$ & $U(-1.00,1.00)$ & $-0.24_{-0.04}^{+0.04}$ & $-0.41_{-0.11}^{+0.11}$ & Dependence of $\gammadm$ on $\log{\msps}$ \\
$\xi_{\gamma}$ & $-0.34$ & $U(-1.00,1.00)$ & $-0.25_{-0.09}^{+0.10}$ & $-0.24_{-0.19}^{+0.23}$ & Dependence of $\gammadm$ on galaxy size \\
$\sigma_{\gamma}$ & $0.06$ & $U(0.02,0.50)$ & $0.051_{-0.019}^{+0.022}$ & $0.07_{-0.03}^{+0.06}$ & Intrinsic scatter in $\gammadm$ \\
$\log{\alpha_{\mathrm{sps}}}$ & $0.10$ & $U(0.00,0.25)$ & $0.101_{-0.008}^{+0.008}$ & $0.148_{-0.019}^{+0.018}$ & Log of the stellar population synthesis mismatch \\
 & & & & & parameter \\\end{tabular}
\end{table*}

\subsection{Dependence on the data used}\label{ssec:whichdata}

The results presented so far are based on fits to image positions and radial magnification ratios of the lenses.
The fitting procedure is meant to simulate a situation in which high-resolution imaging data is available for every lens, from which the radial magnification ratios can be obtained. 
However, when only ground-based imaging data are available, it is not possible to measure radial magnifications, because the strongly lensed arcs are typically not resolved.
In this section we investigate how the constraining power of a sample of 1000 lenses changes in such a case.

We repeated the analysis without using any radial magnification information, that is removing the term relative to $\rmur$ from the likelihood in \Eref{eq:2dintegral}, both for the base and the extended models.
The posterior probability distributions of the two inferences are shown as purple filled contours in \Fref{fig:cp1} and \Fref{fig:cp2} and summarised in \Tref{tab:base_inference} and \Tref{tab:extended_inference}.

With the base model, a fit to image position information alone produces a highly biased result. 
Removing radial magnification information does not appear to produce a decrease in precision: the uncertainty on the hyper-parameters is comparable to that attained in the fit to the whole dataset.
However, a closer look at the posterior probability distribution reveals that the inference on the average dark matter slope parameter, $\mu_\gamma$, is driven by the prior: The values preferred by the data are very close to the lower bound.
Presumably, a less restrictive prior on $\mu_\gamma$ would have resulted in a higher overall uncertainty, and possibly an even more biased inference.

By comparing the results of the fit of the base and extended models to the full dataset, we see that models that are not sufficiently flexible lead to biased inferences.
This last test shows, additionally, that the amount of bias increases as the data become less constraining, at least when working with lens samples with similar properties to the mock that we generated.

Fitting the extended model to image positions only (purple contours in \Fref{fig:cp2}) appears to produce a more accurate answer compared to the base model case: for example, the inferred value of $\asps$ is less than $3\sigma$ away from the truth. However, there is now a strong degeneracy between the three key parameters of the model: the average dark matter mass, the average dark matter slope, and the stellar-population-synthesis mismatch parameter.
We therefore conclude that, in order to disentangle the stellar and dark matter contribution to the total mass of a sample of 1000 strong lenses using only strong lensing data, magnification information is necessary.

\section{Discussion}\label{sect:discuss}

With the experiments presented so far, we quantified the precision and accuracy that can be achieved on the measurement of the distribution of the dark matter density profile and of the stellar mass-to-light ratio of galaxies by statistically combining a sample of 1000 strong lenses.
An important assumption on which our analysis is based is that the source position distribution, the term $\mathcal{B}$ in \Eref{eq:popdist}, is known exactly when making the inference.
We discuss the impact of this assumption in section \ref{ssec:betaprior}.
Subsequently, in section \ref{ssec:pptests} we describe a general strategy with which to decide whether or not a model is sufficiently flexible to fit the data.
In section \ref{ssec:systematics} we discuss possible systematic effects that were not explored by our experiment but that could potentially lead to biases in the inference.
In section \ref{ssec:emulation} we discuss the limitations of our treatment of the lens modelling step.
Finally, in section \ref{ssec:practical} we discuss what steps need to be taken in order to successfully apply our analysis method to a real sample of lenses.

\subsection{The importance of the source position prior}\label{ssec:betaprior}

As discussed above, assuming that the source position distribution is known is equivalent to knowing the strong lensing detection efficiency exactly. This is not a realistic assumption: the process of lens finding consists of several steps, typically including  human visual inspection, which introduces selection effects that are difficult to model from first principles.
In this section we investigate how critical this assumption is for the accuracy of the inference.

\begin{figure}
\includegraphics[width=\columnwidth]{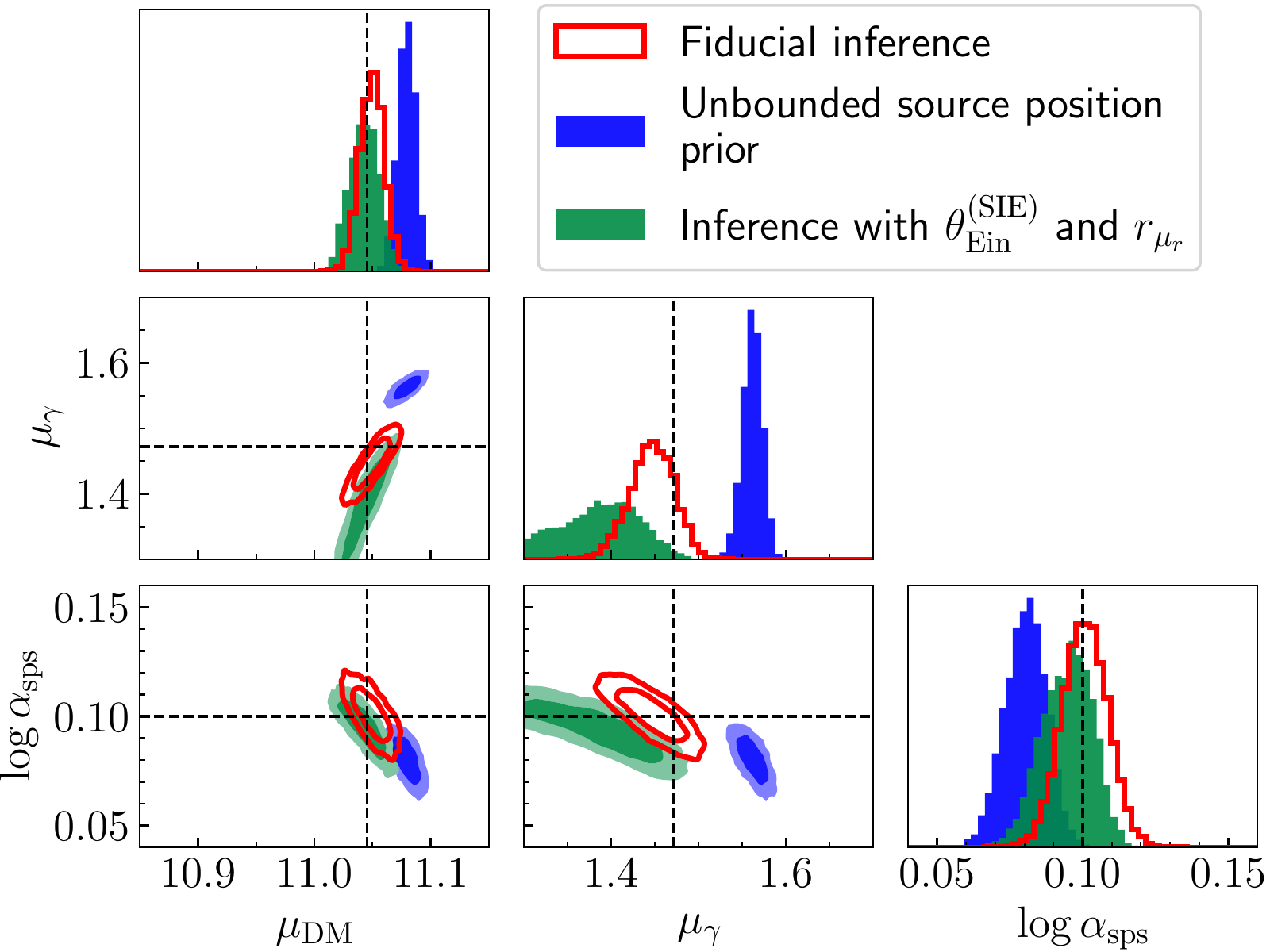}
\caption{
Posterior probability distribution of the hyper-parameters $\mu_{h,0}$, $\mu_{\gamma,0}$, and $\log{\asps}$ obtained under the assumption that source positions are drawn from a uniform distribution in the source plane with no boundary (blue contours) compared to the fiducial inference described in section \ref{ssec:extended} (red contours).
\label{fig:betaprior}
}
\end{figure}

We fitted a modified version of the extended model to the data, in which we adopted an apparently uninformative prior on the source position: We set the model parameter $\betamax$ to infinity in \Eref{eq:betadist}.
This is equivalent to assuming that the sources are drawn from a uniform distribution in the source plane, with no boundary.
The inference on the hyper-parameters describing the average dark matter profile and the stellar population synthesis mismatch parameters are shown in \Fref{fig:betaprior} as blue contours, along with the inference obtained when the prior on the source position is known exactly.
There is a $-0.02$~dex shift in the inference of $\log{\asps}$, which is larger than the uncertainty on that hyper-parameter. The shift on the inference of the average dark matter slope is even bigger in relation to the corresponding uncertainty.

We therefore conclude that, at the precision level afforded by a sample of 1000 lenses, the choice of the source position prior does affect the inference.
This is an example of how a seemingly minor detail, such as modifying the term $\mathcal{B}$ in \Eref{eq:popdist}, can have a sizeable impact on the inference, because the posterior probability distribution depends on the product of a thousand such terms.
This is an important issue that needs to be addressed when analysing a real sample of lenses, either by working with a sample for which the lens detection probability is well characterised or by developing a method that allows one to infer it directly from the data.

Alternatively, we can avoid modelling the source position distribution by compressing the image position information into a model-independent quantity, such as the Einstein radius. 
For example, the half-separation between the two images is a good proxy for the Einstein radius; it is exactly equal to the Einstein radius of a singular isothermal sphere lens:
\begin{equation}\label{eq:tsis}
\tsis = \frac{\theta_1 - \theta_2}{2}.
\end{equation}
Assuming that $\tsis$ approximates the true Einstein radius of a lens well, we can use it as an observable constraint in place of $(\toneobs,\ttwoobs)$.
By doing so, the source position no longer enters the problem explicitly: A derivation similar to that of section \ref{ssec:inferencetech} and Appendix~\ref{sect:appendixa} produces the following expression for the likelihood of observing the data relative to one lens,
\begin{equation}\label{eq:tsisintegral}
\begin{split}
\pr(\datai|\hyperpars) = & \int d\gammadm \int d\log{\mfive} \left\lvert\frac{d\log{\mstartrue}}{d\tein}\right\rvert_{\mstartrue=\mstartrueein} \\
& \pr(\rmurobs|\gammadm,\mfive,\reff,\mstartrueein) \\
& \pr\left(\mobs|\mstartrueein,\asps\right) \\
& \pr\left(\mstartrueein,\reff,\mfive,\gammadm|\hyperpars\right).
\end{split}
\end{equation}
In the integral above, $\mstartrueein$ is now the stellar mass needed to produce a total projected mass within the Einstein radius equal to $\tsis$, as a function of $\mfive$ and $\gammadm$.

By using $\tsis$ in place of $(\toneobs,\ttwoobs)$ we are discarding part of the available information: We no longer fit the distribution in image configuration asymmetry $\asymm$, which is sensitive to the density profile of the lenses. For this reason, we expect the resulting inference to be less precise.
We performed such a fit to $\tsis$ and $\rmur$, the posterior probability distribution of which is shown in green in \Fref{fig:betaprior}.
As expected, the inference is less precise than that  provided by the fiducial analysis.
However, it is more accurate than the case in which an unbounded prior on the source position is assumed.
Compressing the available information into model-independent observables is then a possible way of trading precision for accuracy in the case where it is not possible to obtain an accurate description of the source position distribution.

\subsection{Model selection with posterior prediction}\label{ssec:pptests}

An apparent weakness in our approach is the decision process that led to the extension of the model of section \ref{ssec:extended}: We implemented the extended model after noticing that the base model was unable to recover the truth and stopped improving it once we realised that the new model afforded an accurate inference. This is something that can only be done if we already know the properties of the lens population in detail.
Nevertheless, it is possible to gauge the degree of accuracy of a model by examining its goodness of fit.

When working with Bayesian hierarchical models, goodness of fit is determined with posterior predictive tests: mock observations are generated from the model and these are then compared to selected aspects of the observed data.
In our case, the data consists of a distribution of image positions, image magnification ratios, stellar masses, and half-light radii.
As an example, we show in this section a posterior predictive test that focuses on image positions.

We start by compressing the data into a handful of summary statistics, which we use as quantities to test our model against.
We first reduce the image position distribution to a one-dimensional one by considering the half-separation between images defined in \Eref{eq:tsis}. 
We then consider the mean and standard deviation of the $\tsis$ distribution, $\meantsis$ and $\stdtsis$.
The goal of our posterior predictive test is to determine how likely it is for our model to produce samples with values of these test quantities that are more extreme than the observed ones.

\begin{figure}
\includegraphics[width=\columnwidth]{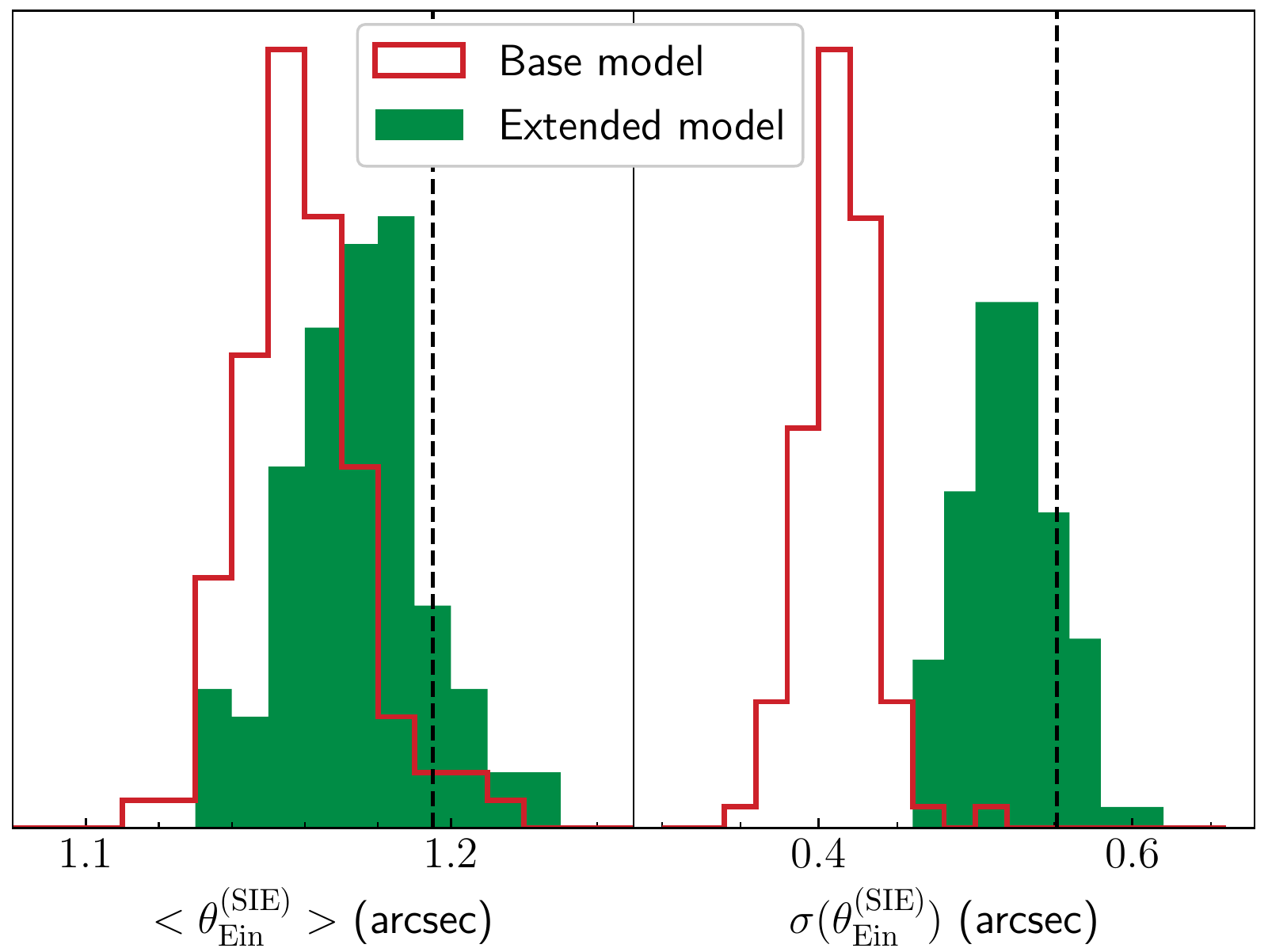}
\caption{
Posterior predicted distribution in the mean value of $\tsis$ (left panel) and in the standard deviation of $\tsis$ (right panel) on samples of 1000 lenses for the base model (red histogram) and extended model inference (green histogram).
The dashed lines indicate the values of $\tsis$ and $\stdtsis$ measured in the observed lens sample to which both models were fitted.
\label{fig:pptsis}
}
\end{figure}

We obtained the posterior predicted test quantities as follows. 
We randomly drew 100 samples from the MCMC of the inference, we generated a sample of 1000 lenses for each draw, measured the value of $\tsis$ of each lens, and finally computed $\meantsis$ and $\stdtsis$ of the sample corresponding to each posterior draw.
The resulting posterior predicted distribution of $\meantsis$ and $\stdtsis$ is shown in \Fref{fig:pptsis}.

The posterior predicted average $\tsis$ obtained from the base model (red histogram) tends to be smaller than the observed value, but realisations in which $\tsis$ is larger are not uncommon.
However, at the same time,  all posterior predicted lens samples have a standard deviation in $\tsis$ that is smaller than the observed one. This means that, if the base model is a faithful description of the truth, it will be extremely unlikely to find a sample of 1000 lenses with a value of $\stdtsis$ as large as the observed one.
We therefore conclude from this test that the base model is unable to reproduce the observed distribution in Einstein radius of the lens sample in detail.

This test on its own tells us that the base model does not provide a good fit, but does not give explicit indications as to how to improve it.
Additional posterior predictive tests can provide further insight: for example, the posterior predicted lens samples are also unable to match the observed correlations of $\tsis$ with $\msps$ and $\reff$, which suggests that correlations between the dark matter distribution and the structural parameters of the galaxy might be needed to provide a good description of the sample.
The extended model introduced in section \ref{ssec:extended} allows for such correlations and provides a much better match between its posterior predicted Einstein radius distribution and the observed one, as shown by the green histograms in \Fref{fig:pptsis}.

In general, the choice of the test quantity is arbitrary and must reflect the aspect of the model accuracy that the user wishes to assess, depending on their science goal. 
We are mostly interested in using our model to capture the average properties of the lens population; therefore, we focused on the mean and standard deviation of $\tsis$, but in principle other choices are possible. 
For example, to check whether the model is able to reproduce the presence of outliers, one can use a high percentile of the $\tsis$ distribution as a test quantity.

In summary, posterior predictive tests provide a way of assessing the goodness of fit of a Bayesian hierarchical model and can be used to improve an existing model or discriminate between alternative ones.
Nevertheless, we stress that these tests are by no means a way of building a model purely on the basis of the available data: physical insight should always be the guiding principle of any astrophysical model.

\subsection{Possible sources of systematic errors}\label{ssec:systematics}

The method presented in this work produces an accurate inference of $\asps$ and the dark matter distribution when our extended model is fitted to a population of lenses with the same properties as the mock that we generated for our experiment.
However, there could be scenarios in which the same model returns a biased answer: Any discrepancy between the model being fitted and the truth underlying the data is a potential source of bias.
In our experiment we focused on what we consider to be the most important unknown in strong lens modelling: the dark matter density profile. Our mock lenses were generated with a non-trivial prescription for determining the dark matter distribution.

In principle, we could increase the complexity of the mock even further, for example by allowing for a bimodality in the dark matter content. Massive galaxies are known to have a bimodal distribution in their inner surface brightness profile \citep[cored or cuspy; see e.g.][]{Lau++06} and in their velocity structure \citep[the slow- and fast-rotator dichotomy: see][ and references therein]{Gra++19}, and the lensing and kinematics study of \citet{O+A18} suggest the presence of a bimodal distribution in the inner slope of the dark matter halo.

Moreover, our mock was generated by assuming a single value for the stellar population synthesis mismatch parameter of the whole sample. In reality, there could be variations in the value of $\asps$ among the population \citep{Tre++10, Cap++12, C+v12, Son++15} and even spatial variations within individual objects \citep{vDo++17, Son++18b}.
Redshift evolution of either the dark matter distribution or $\asps$ is also something that might occur in reality but is not considered in our mock.

Finally, the line-of-sight structure also contributes to the observed lensing signal.
The lensing effect of the line-of-sight structure can be modelled ---at the scale of a single lens--- as a constant sheet of mass with surface mass density $|\kappa_{ext}| \lesssim 0.1$ \citep[see e.g.][]{Mil++20}.
This external mass sheet is degenerate with the dark matter halo of the lens: For example, a positive value of $\kappa_{ext}$ can mimic the effect of increasing the dark matter mass enclosed within the Einstein radius and making the dark matter density profile shallower.
However, if the line-of-sight structure correlates with the stellar distribution, neglecting it can potentially lead to biases on the inference of $\asps$ as well.

In order to quantify the impact that these possible additional levels of complexity in the true mass structure of lens galaxies can have on the accuracy of the inference, it is necessary to test the method on dedicated simulations. However, this is beyond the goals of the present work.

\subsection{Limitations of the lens modelling emulation}\label{ssec:emulation}

In order to reduce the computational burden of our investigation, we emulated the lens modelling step: We assumed that the information content of the strong lensing data of each lens can be summarised with two image positions and a radial magnification ratio.
The rationale for the use of the radial magnification ratio is that this quantity is related directly to the radial density profile of the lens and can be measured once the width of the arc and the counter-image are obtained.

On the one hand, this is a conservative choice: In principle, the elongation and curvature of the arcs can also be used to constrain a lens model \citep[although in practice part of this information is needed to determine the azimuthal structure of the lens: see][]{Bir21}.
On the other hand, when generating the mock we added an observational error of $0.05$ to $\rmur$. 
This value was estimated on the basis of the typical uncertainty on the radial density profile of a lens obtained by \citet{Sha++21} when modelling high-resolution images of lenses, under the assumption that the constraining power comes primarily from the measurement of the radial magnification ratio \citep[][ also made this assumption in the interpretation of their measurements]{Sha++21}.
However, a small uncertainty on the radial profile can also be the result of fitting a model that has an overly simple azimuthal structure: in such cases, the model is over-constrained and the uncertainty on the radial profile is underestimated \citep{Koc21}.
Determining what model-independent quantities can be measured with different aspects of the data is an important problem, but is beyond the scope of this work.
Nevertheless, there is the possibility that the true uncertainty on the radial magnification ratio that can be obtained in practice is higher than the value assumed in our experiment. In that case, our predicted uncertainty on the model hyper-parameters will be underestimated.

\subsection{Application to real samples of lenses}\label{ssec:practical}

The inference method presented in this work consists in fitting a model describing the population of lenses directly to the full ensemble of imaging data of a large sample of lenses with a Bayesian hierarchical approach.
This method has never been applied to a real sample of lenses. 
In our experiment we simplified the problem by emulating the lens modelling step, compressing the strong lensing data of each lens down to three numbers.
In reality, the data consist of images made of thousands of pixels.
In order to fit these data, it is necessary to model the full surface brightness distribution of the lens and of the background source, and, because real lenses are not axisymmetric, to allow for additional degrees of freedom in the mass model related to the azimuthal structure. 

Currently available modelling codes are able to deal with these complexities \citep[see e.g.][]{B+A18, Nig++19}: The background source can be reconstructed with a pixellated model and the lens can be modelled as the sum of elliptical mass components.
In practice, however, the lens modelling step needs to be automated, as the traditional approach requires a lot of human interaction, an approach that does not scale well to samples of thousands of lenses.
Recently, there has been progress on this front: \citet{NDM18} developed an automated lens-modelling algorithm and showed it to be accurate in a variety of cases.
Machine learning can also be used to perform fast automated lens parameter inferences \citep[see e.g.][]{HPM17, Chi++20, Sch++21, Par++21}: In particular, \citet{Wag++21} showed how it is possible to carry out hierarchical inferences on lens populations with Bayesian neural networks \citep{CPL21}.
However, it is not clear whether or not these methods are able to sample the posterior probability distribution of the lens parameters in a way that is sufficiently accurate for our purposes: dedicated tests are needed.

On a related issue, our set of assumptions enabled us to greatly simplify an otherwise very computationally intensive step in our analysis: the marginalisation over the parameters describing individual lenses, \Eref{eq:4dintegral}.
In principle, to compute the likelihood of each set of values of the hyper-parameters, one must average over all possible values taken by many individual lens parameters. In a sample of real lenses, these are at the very least the four parameters already employed in our model, plus additional ones describing the azimuthal structure of each lens and the surface brightness distribution of the source.
Moreover, there is the added burden that the data vector is an image instead of a handful of numbers.

Clearly, it is necessary to find a way to approximate the computation of the integrals of the kind of \Eref{eq:4dintegral} in practice.
The method that we used, namely spline integration on a grid, does not scale well to a higher number of dimensions.
One of the most commonly used approaches to compute fast integrals is Monte Carlo integration paired with importance sampling, but that method can lead to biases in cases in which the samples used for the integration do not cover the integrand function well over its entire support.
We therefore leave this as a major open computational issue.

One could argue that the marginalisation over the individual lens parameters is not a necessary step in a Bayesian hierarchical analysis:
The posterior probability distribution of the full ensemble of parameters, both those describing the population and the individual lens ones, can be explored with a Gibbs sampling approach.
While that is true in principle, Gibbs sampling fails to converge in a regime where the individual object parameters are under-constrained by the data, which is the case when fitting complex mass models to strong lensing data, rendering such an approach impractical.

Finally, while all the lenses and sources  in our mock are at the same redshift, this is not true in real samples of lenses.
Varying the source redshift at fixed lens properties changes the Einstein radius.
This means that having a distribution of source redshifts can allow us to probe the mass of the lens at different physical apertures. In principle, this information can be used to better constrain the lens structure: For example, measuring how the Einstein radius varies as a function of source redshift at fixed stellar mass, stellar density profile, and lens redshift can tell us about the dark matter halo density profile. However, in practice this signal is swamped by the scatter in the lens population, both intrinsic and observational (on $\msps$). Therefore it is unclear whether working with a distribution of source redshifts can actually improve the inference.


\section{Conclusions}\label{sect:concl}

We present a Bayesian hierarchical inference method for statistically combining strong lensing constraints from a large sample of lenses with the goal of measuring key aspects of the inner structure of lens galaxies: the stellar mass-to-light ratio, the dark matter mass, and the dark matter density profile.
We tested the method on a simulated sample of 1000 lenses generated under the simplifying assumption that all lenses are axisymmetric and all lensed sources are point-like.
We fitted two models to the mock observations, with increasing degrees of complexity. In both cases, the functional form of the fitted model was different from the properties of the simulation, both in terms of the density profile of individual lenses and in terms of the population distribution of the dark matter halo parameters.
We found the following:
\begin{itemize}
\item When image position and magnification information are used to constrain the model, a sample of 1000 lenses can constrain the stellar population synthesis mismatch parameter, the dark matter normalisation, and the inner slope with very high precision and accuracy compared to current observations.
This means that it is possible to calibrate stellar mass measurements with high accuracy and obtain a firm detection of the effect of baryonic contraction on the dark matter halos, and therefore to settle the dark matter core versus cusp debate at the halo masses characteristic of galaxy-scale strong lenses.
\item In order to obtain an accurate inference, the model describing the population of lenses must allow for correlations between the parameters of the dark matter component and all dynamically relevant properties of the lens galaxies, such as the stellar mass and half-light radius.
\item When fitting image positions only, it is still possible to obtain an accurate inference, but by paying a large cost in terms of precision: even with 1000 lenses we cannot break the degeneracy between the dark matter profile and the stellar-mass-to-light ratio. Complementary information from another dynamical probe ---such as weak lensing--- is needed in that case.
\item A necessary condition for obtaining an accurate inference is being able to provide a faithful description of the source position probability distribution or, equivalently, to know the detection probability of a lens as a function of its image configuration.
Alternatively, fitting the Einstein radius instead of the image positions provides a way of maintaining accuracy, but at the cost of precision.
\item Posterior predictive tests allow one to evaluate the goodness of fit of a Bayesian hierarchical inference and are therefore a useful tool for building accurate models.
\end{itemize}
The tests carried out in this paper provide a first forecast of the potential constraints that large samples of strong lenses can provide.
In order to implement the method in practice, several challenges still need to be addressed. These include measuring the redshifts of large numbers of lenses and relative sources, making the individual lens modelling step as automated as possible, and ensuring that the likelihood evaluation and the marginalisation over the many parameters describing individual lenses, a requirement of our method, can be carried out in an accurate and computationally sustainable way.

This work was the first of a series. In a second paper we will quantify the constraining power of a combination of image position and time-delay information, and in a third paper we will use the number density of a complete sample of lenses as an additional constraint.


\begin{acknowledgements}

We thank Phil Marshall for useful discussions and suggestions.
AS acknowledges funding from the European Union's Horizon 2020 research and innovation programme under grant agreement 792916 (Halos2020).
MC acknowledges support by the EU Horizon 2020 research and innovation programme under a Marie Sk{\l}odowska-Curie grant agreement 794474 (DancingGalaxies).

\end{acknowledgements}

\bibliographystyle{aa}
\bibliography{references}

\appendix
\section{Marginalisation over the stellar mass and source position}\label{sect:appendixa}

In order to evaluate the posterior probability distribution of the model hyper-parameters given the data, we need to compute integrals of the kind of that in \Eref{eq:4dintegral}.
Let us consider the first term of the integrand function. This is the product of four terms, one for each observable:
\begin{equation}
\begin{split}
\pr\left(\datai|\mstartrue,\asps,\reff,\mfive,\gammadm,\beta\right) = & \pr(\toneobs|\mstartrue,\reff,\mfive,\gammadm,\beta) \pr(\ttwoobs|\mstartrue,\reff,\mfive,\gammadm,\beta) \times \\
& \pr(\rmurobs|\mstartrue,\reff,\mfive,\gammadm,\beta) \pr(\mobs|\mstartrue,\asps).
\end{split}
\end{equation}
Because the two image positions are measured exactly, each of the first two terms is a Dirac delta function,
\begin{equation}
\pr(\toneobs|\mstartrue,\reff,\mfive,\gammadm,\beta) = \delta(\theta_1(\mstartrue,\reff,\mfive,\gammadm,\beta) - \toneobs),
\end{equation}
and a similar expression holds for the term relative to image $2$.
Here $\theta_1$ indicates the position of image 1 as predicted by the model parameters and is a function of the latter.
In order to integrate out these Dirac delta functions, we first apply the following variable change:
\begin{equation}
(\log{\mstartrue},\beta) \rightarrow (\theta_1,\theta_2).
\end{equation}
If $\detJ$ is the Jacobian determinant of this variable change, \Eref{eq:4dintegral} then becomes
\begin{equation}
\begin{split}
\pr(\datai|\hyperpars) = & \int d\gammadm \int d\log{\mfive} \iint d\theta_1 d\theta_2 |\detJ| \delta(\theta_1 - \toneobs)\delta(\theta_2 - \ttwoobs) \times \\
& \pr(\rmurobs|\mstartrue(\theta_1,\theta_2),\reff,\mfive,\gammadm,\beta(\theta_1,\theta_2)) \pr(\mobs|\mstartrue(\theta_1,\theta_2),\asps) \\
 & \pr\left(\mstartrue(\theta_1,\theta_2),\reff,\mfive,\gammadm,\beta(\theta_1,\theta_2)|\hyperpars\right).
\end{split}
\end{equation}
We can now integrate over $\theta_1$ and $\theta_2$ to obtain
\begin{equation}
\begin{split}
\pr(\datai|\hyperpars) = & \int d\gammadm \int d\log{\mfive} \left\lvert\detJ\right\rvert_{(\mstartrue,\beta)=(\mstartrueein,\betaein)}\\
& \pr(\rmurobs|\gammadm,\mfive,\reff,\mstartrueein,\betaein) \\
& \pr\left(\mobs|\mstartrueein,\asps\right) \\
& \pr\left(\mstartrueein,\reff,\mfive,\gammadm,\betaein|\hyperpars\right),
\end{split}
\end{equation}
where we define $\mstartrueein$ and $\betaein$ as the values of the true stellar mass and source position needed to produce images at $\toneobs$ and $\ttwoobs$.
We point out that, for certain combinations of values of the lens model parameters, the source is not strongly lensed, and therefore $\theta_2$ is not defined. In those regions of the parameter space, the likelihood is simply zero.
\end{document}